\documentclass[aps,physrev,twocolumn,superscriptaddress,preprintnumbers,showkeys]{revtex4-2}

\usepackage{amsmath,amsthm,amssymb}
\usepackage{appendix}
\usepackage{mathtools}

\newcommand{\braket}[1]{\langle #1 \rangle}
\newcommand{\avg}[1]{\langle #1 \rangle}

\begin{document}


\title{Detecting the Axion-Photon Conversion Background}


\author{Felix Weber}
\email{fweber@caltech.edu}
\affiliation{Cahill Center for Astronomy and Astrophysics, MC 249-17 California Institute of Technology, Pasadena CA 91125, USA}

\author{Vikram Ravi}
\affiliation{Cahill Center for Astronomy and Astrophysics, MC 249-17 California Institute of Technology, Pasadena CA 91125, USA}


\date{\today}

\begin{abstract}
The potential to detect axion dark matter through astrophysical processes has shown high promise in recent years. We therefore expand on previous work studying the axion-to-photon conversion efficacy of neutron stars and the interstellar medium (ISM) in this endeavor, respectively. For neutron stars (NS), we examine the possibility of a background signal emanating from all NS magnetospheres in the galaxy. Using a heuristic Galactic model, we find a significant background signal emanating from such magnetospheres in the Milky Way. This signal, while weak in absolute power ($\gtrsim 1$\,mJy\,sr$^{-1}$ from the Galactic Center, at 2 GHz), can be detected through new statistical techniques with current instrumentation like the Atacama Large Millimeter Array (ALMA) at high radio frequencies (200--950\,GHz). These techniques make use of higher order statistics like spectrally-limited ($\sim 300$\,km\,s$^{-1}$) increases in confusion noise levels and kurtoses of survey images, and also show promise for general population estimation techniques. \\
\indent For the ISM, we consider Primakoff processes between free electrons and axions, and derive typical signal strengths of $10^{-15}$\,Jy\,sr$^{-1}$ $\cdot$ $m_a$/eV, with a local, cosmological upper bound of $10^{-8}$\,Jy\,sr$^{-1}$ $\cdot$ $m_a$/eV. Hence, we find that any diffuse axion signal from the ISM and other, large-scale, astrophysical plasmas to be too weak to be detected with modern technologies. We therefore find that the best avenue towards detecting a potential quantum chromodynamics (QCD) axion dark-matter particle is through the radio imaging of large swaths of the Galactic Center and other regions where we expect large numbers of pulsars and neutron stars. 
\end{abstract}

\keywords{Axion --- Pulsars --- Radio Interferometry --- Interstellar Medium}

\maketitle


\section*{Introduction}

\indent Axions have the potential to resolve two important problems in modern physics: the Strong-CP problem and the dark matter problem \cite{Zwicky}, challenging the Standard Model, and our current understanding of the large-scale structure of the Universe and gravitation, respectively. The history of Axions started with Peccei and Quinn proposing in 1977 that CP-symmetry breaking may have changed over time to help explain the current matter/anti-matter imbalance of the Universe, as part of a Strong-CP solution \cite{PQ1977}. Wilczek \cite{AxionPaper1} and Weinberg \cite{AxionPaper2} then showed that the field Peccei and Quinn prescribed in their Strong-CP solution would produce a Goldstone-boson, which they dubbed the axion.

\indent With the addition of the axion particle, there exist different prescriptions for how the field from which the Wilczek-Weinberg QCD axion arises directly interacts with different components of the Standard Model. The most popular of these are the KSVZ and DFSZ models. For a concise treatment of these and other models, refer to \cite{di_Cortona_2016}. The most important general result of these models is a direct coupling to quarks that gives rise to an effective coupling with photons at low energy scales. Hence, it may be possible to indirectly detect axions through the electromagnetic (EM) spectrum. Furthermore, it was shown that at low axion masses, these particles would act precisely as a near-invisible source of matter in the Universe that interacted primarily through the gravitational force at large scales. Today the lower bound on the axion mass is set by the requirement that axions in the early universe do not breach critical density and halt inflation \cite{AxionDownBound}: $\sim 10^{-7}$\,eV. The upper bound of $\sim 10^{-2}$\,eV, on the other hand, is set by known hydrogen fusion rates in stars and supernovae \cite{Buschmann_2022} \cite{Caputo:2024oqc}. The remaining mass range corresponds to EM radiation within the low radio to terahertz frequency range.

\indent Hence, QCD axions represent a \emph{testable} solution to two fundamental problems in modern physics: the dark matter problem and the Strong-CP problem. However, the effective coupling between axions and photons is predicted to be extremely weak. Eq.~\ref{eq:effectiveaxion} gives the effective Langrangian of the axion-photon interaction, from which all current EM searches for axions are derived. 

\begin{equation}
    \begin{split}
    \mathcal{L}_{\alpha \gamma} &= g_{\alpha \gamma\gamma}\frac{\alpha}{8\pi} a F_{\mu \nu} \tilde{F}^{\mu \nu} \\
    g_{\alpha \gamma \gamma} &= \left[0.203(3) \frac{E}{N} - 0.39(1) \right] \frac{m_a}{\text{GeV}^2}
    \end{split} \label{eq:effectiveaxion}
\end{equation}

\indent The effective theory is only dependent on two parameters: $m_a$ and $E/N$. On a fundamental level, $m_a$ is a function of quark and pion masses and the fundamental decay constant $f_a$ that parametrizes the potential of the scalar quantum field, $V(\varphi)$, that Peccei and Quinn introduced. On the other hand, the $E/N$ parameter is model-dependent and reflects how the Standard Model is modified by the axion. For example, for KSVZ and DFSZ, respectively, $E/N$ evaluates to 0 and $8/3$ \cite{di_Cortona_2016}. This makes KSVZ the easiest model to test for, as the coupling constant is larger in absolute magnitude. That being said, one should not presume a specific $E/N$ in the most general case. Hence, it is most common to search for axions over a space defined by $m_a$ and $g_{\alpha \gamma \gamma}$. 

\indent Nearly all experiments make use of variations on a process known as the Primakoff effect, whereby an axion converts to a photon (or vice-versa) via an interaction with virtual photons \cite{Sikivie}. Subsequently, axion experiments can be grouped into two groups: those that attempt to directly convert the local axion field into an EM signal, and those that seek to detect the associated EM signal from elsewhere in the Universe. Historically, the former was preferred, however even modern experiments (e.g., the CERN Axion Solar Telescope; CAST \cite{CAST}) have struggled to reach sensitivities required for current coupling theories. Hence, in the last decade, astrophysical searches have become more popular, owing to the abundance of extreme phenomena in the Universe that beat even the most ideal laboratory conditions. It is this approach that we consider, expanding on work by Berghaus et al. \cite{dsa_std_mod} for neutron stars, and Kelley and Quinn \cite{Kelley_2017} for the interstellar medium (ISM). 

\indent While \cite{Kelley_2017} argues that the ISM of our own galaxy would produce a detectable radio signal, their core calculation assumes a standard turbulent cascade of the ISM magnetic field, down to the radio wavelength ($\sim$ 1 m$^{-1}$) required for resonant conversion. The turbulent spectrum of the ISM is not well understood, but solar astronomers have been able to study the plasma turbulence in the solar magnetosphere. \cite{Alexandrova_2012} has shown that below the electron scale, $\rho_e$, given by the electron's Larmor radius, the magnetic power spectrum follows an exponential cutoff, $k^{-8/3} \exp(-k\rho_e)$. For the electron scale to be in the radio wavelength, and hence for the process described in \cite{Kelley_2017} to be significant, the local average magnetic field strength in the ISM must be $\sim$20 mG. This is over four orders of magnitude larger than that expected for the ISM \cite{Han_2007}. For more typical few-$\mu$G fields, the electron scale will be $\sim 10$ km, resulting in a reduced magnetic power spectrum, and hence a reduced axion signal. Relative to the standard $k^{-8/3}$ cascade this reduction will be by over a hundred orders of magnitude at radio wavelengths due to the exponential decay term. We therefore will demonstrate in Section \ref{sec:axioninism} a different potential, but weak, axion signal process in the ISM that does not rely on turbulent assumptions: the direct conversion of axions to photons by interactions with free charged particles in ISM plasmas, rather than any residual magnetic fields. 

\indent On the other hand, neutron stars represent some of the most extreme environments observable to us today. With the extreme EM fields present in most neutron stars, Berghaus et al. \cite{dsa_std_mod} and many others before them (\cite{Pshirkov_2009}, \cite{Safdi_2019}, \cite{Leroy_2020}) have explored the use of resonant `halos' around these stars where infalling axions quickly convert to photons in the stars' magnetospheres. Much work has been done on understanding this process on an individual level, from pulse profile simulations \cite{Battye_2021} to related individual axion-pulse searches \cite{2023PhRvD.108f3001B}. Based on known population statistics, Berghaus et al. \cite{dsa_std_mod} have shown that there \emph{may} exist a neutron star somewhere in the Milky Way, yet to be discovered, that would produce a spectral line strong enough to be detected with modern radio instrumentation. To add to their and others' work, we will demonstrate in Section \ref{sec:axionbackground} through direct population synthesis, how all pulsars, in harmony, would produce a diffuse (max. $\sim 1$ mJy/str for high $m_a$), but stochastic, background signal in the sky that would be detectable under higher-order statistical tests (e.g., increases in confusion noise) for existing radio instrumentation. 

\section{Background from Galactic Pulsar Magnetospheres\label{sec:axionbackground}}

\indent Radio astronomy instrumentation is amidst a watershed moment, with the advent of ALMA at millimeter wavelengths, and the Deep Synoptic Array (DSA) and the SKA at MHz to GHz radio frequencies. These instruments allow for efficient surveys with large collecting areas\footnote{By order of magnitude: DSA will have a collecting area of $\approx 5 \times 10^4 \text{ m}^2$ across all antennae. Within its operational range, it will have $T_{\text{SEFD}} \cdot \sqrt{\Delta B \Delta T} \approx 0.2 \text{ K}$ for a purely flat, diffuse backgorund when including autocorrelation terms. For an hour-long observation of a spectral line over 500 kHz, this corresponds to an average sensitivity of 4 $\mu$K.} suited for faint, diffuse spectral-line signals originating from axions and other sources. 

\indent To properly calculate the axion EM background associated with neutron stars, we require an accurate prescription of the power from the signal, and an accurate neutron-star population model. Again, \cite{dsa_std_mod} provides such a prescription in their Eq. 2.11 (provided as Eq.~\ref{eq:ax_pls_sig} in this paper for a KSVZ axion model). For the purposes of this paper, the geometric term in Eq.~\ref{eq:ax_pls_sig} is phase and population averaged to $\sim$0.1, assuming random orientations of $\hat{r}$, and $\theta_m = \pi/2$. We will also assume a constant virial velocity $v_0 = 200$ km/s. A neutron-star population model, however, will be created specifically for this task, rather than borrowing from previous models. This is to ensure flexibility in prior distributions of different neutron-star parameters, and tuning the model to fit general characteristics of the Milky Way. 

\begin{equation}
    \begin{split}
    \frac{d\mathcal{P}}{d\Omega} = 3.38 \times 10^{3} \text{ W }  \left(\frac{R_\text{NS}}{10 \text{ km}}\right)^{5/2} \left(\frac{m_a / 2\pi}{ \text{GHz}}\right)^{10/3} \\
    \left(\frac{B_0}{10^{12} \text{ G}}\right)^{5/6} \left(\frac{P}{ \text{sec}}\right)^{7/6} \nonumber 
    \left(\frac{\rho_{\text{DM}}}{0.4 \text{ GeV cm}^{-3}}\right) \\\left(\frac{M_\text{NS}}{M_\odot}\right)^{1/2} \left(\frac{200 \text{ km/s}}{v_0}\right) \cdot \left| 3 \cos(\theta) \hat{m} \cdot \hat{r} - \cos(\theta_m)\right|^{5/6} 
    \end{split}
    \label{eq:ax_pls_sig}
\end{equation}

\indent Note, much work has been done on the subject of Axion `clouds' around pulsars that form from run-away pair production processes converting into Axions \cite{Prabhu_2021}. These clouds have been shown to produce a potentially significant signal once saturated, however they will dissipate once pair production ceases - i.e. once a pulsar ages to the death line. We therefore do not consider Axion clouds on two grounds: first, a galactic background signal will be dominated by large-period, low-$B$ neutron stars that have long gone radio-quiet. These stars no longer host Axion clouds, but their magnetospheres will still be able to convert dark matter Axions. Secondly, Axion clouds are a direct test of the Axion theory itself, but do not reveal if dark matter is itself composed of Axions. We therefore pose our results as an estimate of the Axion line background from dark matter alone: a `worst case' scenario where Axion clouds do not exist. Naturally, if Axions are dark matter, we should expect both a DM and cloud signal - an overall stronger signal - and work will have to be done to disentangle the two. Hence, our results will reveal whether the DM component can be measured directly.

\subsection{The Neutron-Star Population Model}

\indent In creating our own model, we borrow and modify methods primarily from two previous papers. For a spatial distribution mimicking the Milky Way's spiral structure, we use a similar technique as in \cite{PulsarSpirals}, and for appropriate parameter distributions for neutron stars observed as pulsars, we borrow from \cite{PulsarPop} which has been validated against more recent pulsar catalogs. We do this to bridge two important philosophies in neutron-star population modeling: 1) getting accurate spatial predictions of neutron stars in the observable space of the pulsar population as in \cite{PulsarSpirals}, and 2) modeling the full range of physically possible neutron stars, as in \cite{PulsarPop}. For the kinematic evolution of neutron stars, we assume a simple, constant magnetic field ($B$) spin-down model, with an analytic solution:

\begin{align}
    B \sin(\alpha) &\approx 3.2 \times 10^{19} \text{ G } \sqrt{P \dot{P} / \text{sec}} \label{eq:B_approx} \\
    \Rightarrow P(t) &\approx \sqrt{\left(\frac{B \sin(\alpha) t}{ 3.2 \times 10^{19} \text{ G}}\right)^2 + P(0)^2 } \label{eq:p_time}
\end{align}

\indent For now, we will make the ansatz of a perpendicular inclination, such that $\sin(\alpha) =1$, as distributions of $\alpha$ in the current literature are an open question. When we first generate a neutron star at some time $t<0$, the four parameters, $\{P, \rho^2, z, B\}$, are drawn from appropriate distributions, listed in table \ref{tab:pls_mod_params}. Each neutron star is then time-evolved to the present time, $t=0$, using Eq. \ref{eq:p_time}. Note, $\dot{P}$ can be calculated at any time using Eq. \ref{eq:B_approx}. We also make a constant $B$ assumption in our models, ignoring Ohmic dissipation and other magnetic field weakening mechanisms, as well as any spin-up processes that would produce millisecond pulsars.

\begin{table}[h]
    \centering
    \begin{tabular}{||c|c||c|c||c|c||}
        \hline
        $P$ & Log-Uniform & $P_1$ & 2 ms & $P_2$ & 15 ms\\
        $\rho^2$ & $\chi^2$ & $k$ & 2 & $\sigma_\rho$ & 5 kpc \\
        $z$ & Exponential & $z_0$ & 1 kpc & & \\
        $B$ & Log-Normal & $\mu_B$ & $12.5$ [log$_{10}$ G] & $\sigma_B$ & $0.7$ [log$_{10}$ G]\\ \hline
    \end{tabular}
    \caption{Distributions used for initial neutron star parameters at birth. Taken, with some modifications, from the ansätze made in \cite{PulsarPop}.}
    \label{tab:pls_mod_params}
\end{table}

\begin{figure}[h]
    \centering
    \includegraphics[width=0.99\linewidth]{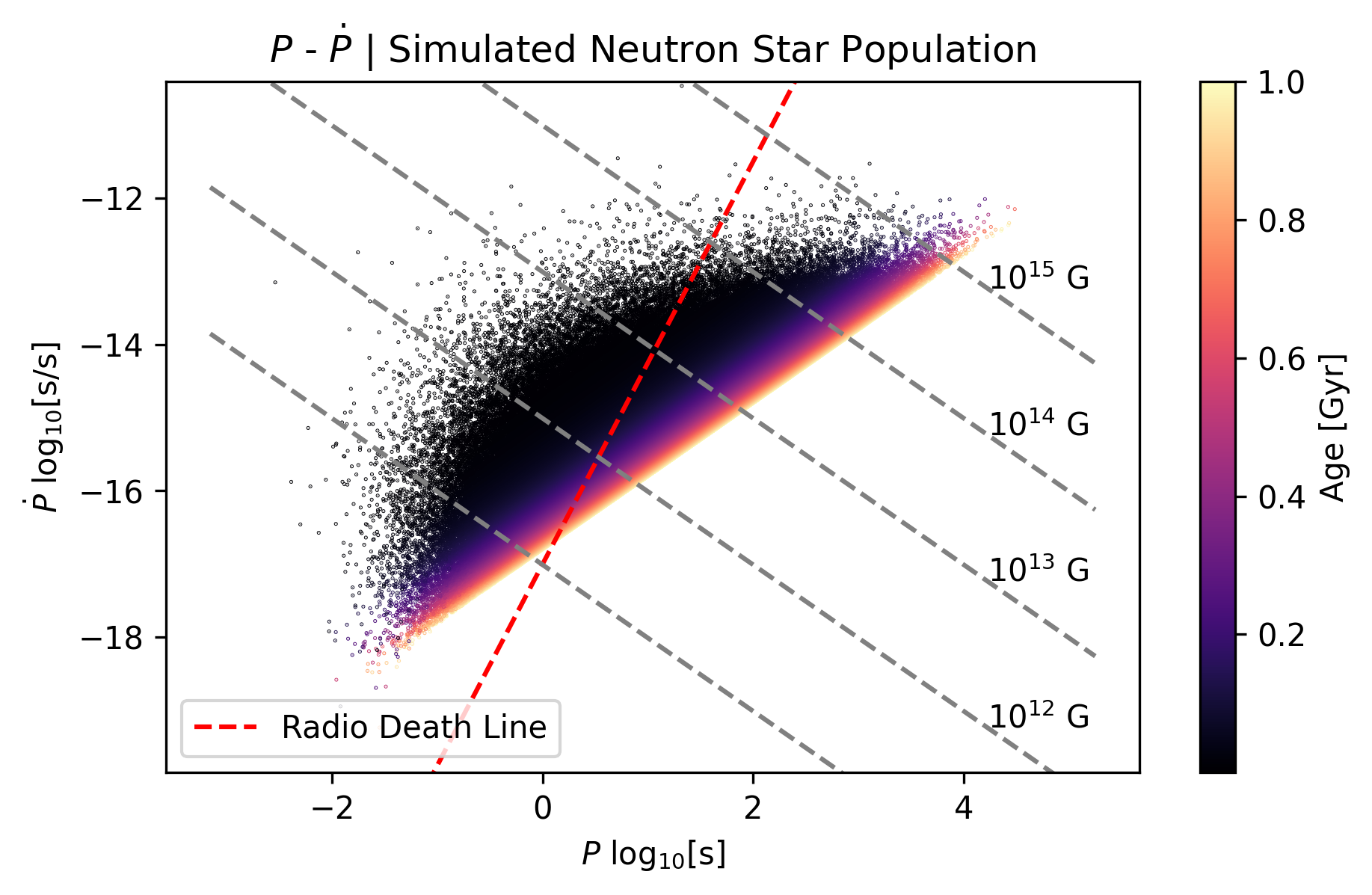}
    \caption{A complete $P$ - $\dot{P}$ diagram of our population model with $10^7$ neutron stars, with objects colored by age. A pulsar death line given by \cite{PulsarDeath} with $\beta \sim 10^{-2}$ to show a general lower bound of expected radio emission is given in red. Magnetic field strengths are labeled in gray.}
    \label{fig:PulsarPopulation}
\end{figure}

\indent Our population starts at some negative $t_0$, which we will take to be $t_0 = -10^9$\,yr. Assuming a constant average neutron star birth rate of $1/T_\text{pls}$, time-steps can be drawn from an exponential distribution with mean $T_\text{pls}$, where every such time-step, a new neutron star is generated until the present time. We take $T_\text{pls} = 100$\,yr, roughly coinciding with the known core-collapse supernova rate of the Milky Way, giving a total population of $10^7$ neutron stars. Figure~\ref{fig:PulsarPopulation} plots all neutron stars in our population model. Note that we include radio-quiet systems - those below the radio death line - in our axion signal calculation, as their magnetic fields will still allow for axion conversion.

\indent Galactic coordinates are then assigned to each neutron star using a similar method as outlined in Sections 3.1.1 and 3.2.1 of \cite{PulsarSpirals}. For each of four major spiral arms, an analytic curve tracing their paths is defined (Eq. \ref{eq:spiral_struct}) using the given parameters in Table 2 of \cite{PulsarSpirals}. Each neutron star is then randomly assigned to one of the four arms and placed on the curve given by their $\rho$-position (Eq. \ref{eq:spiral_struct}). Each arm is then diffused angularly by a normal-random correction term $\Delta \theta$ that exponentially scales smaller as $\rho$ grows. This creates a diffuse bulge in the Galactic Center, while maintaining spiral structures at medium and far distances to the core. 

\begin{align}
    \theta_n(\rho) &= k_n \ln(\rho/\rho_{0,n}) + \theta_{0,n} \label{eq:spiral_struct} \\
    \Delta\theta &= \theta_c \exp(-0.35 \rho/\text{kpc}), \text{ } \theta_c \sim \mathcal{N}(0, 2\pi) \label{eq:th_diffuse}
\end{align} 

\begin{figure}[h]
    \centering
    \includegraphics[width=0.99\linewidth]{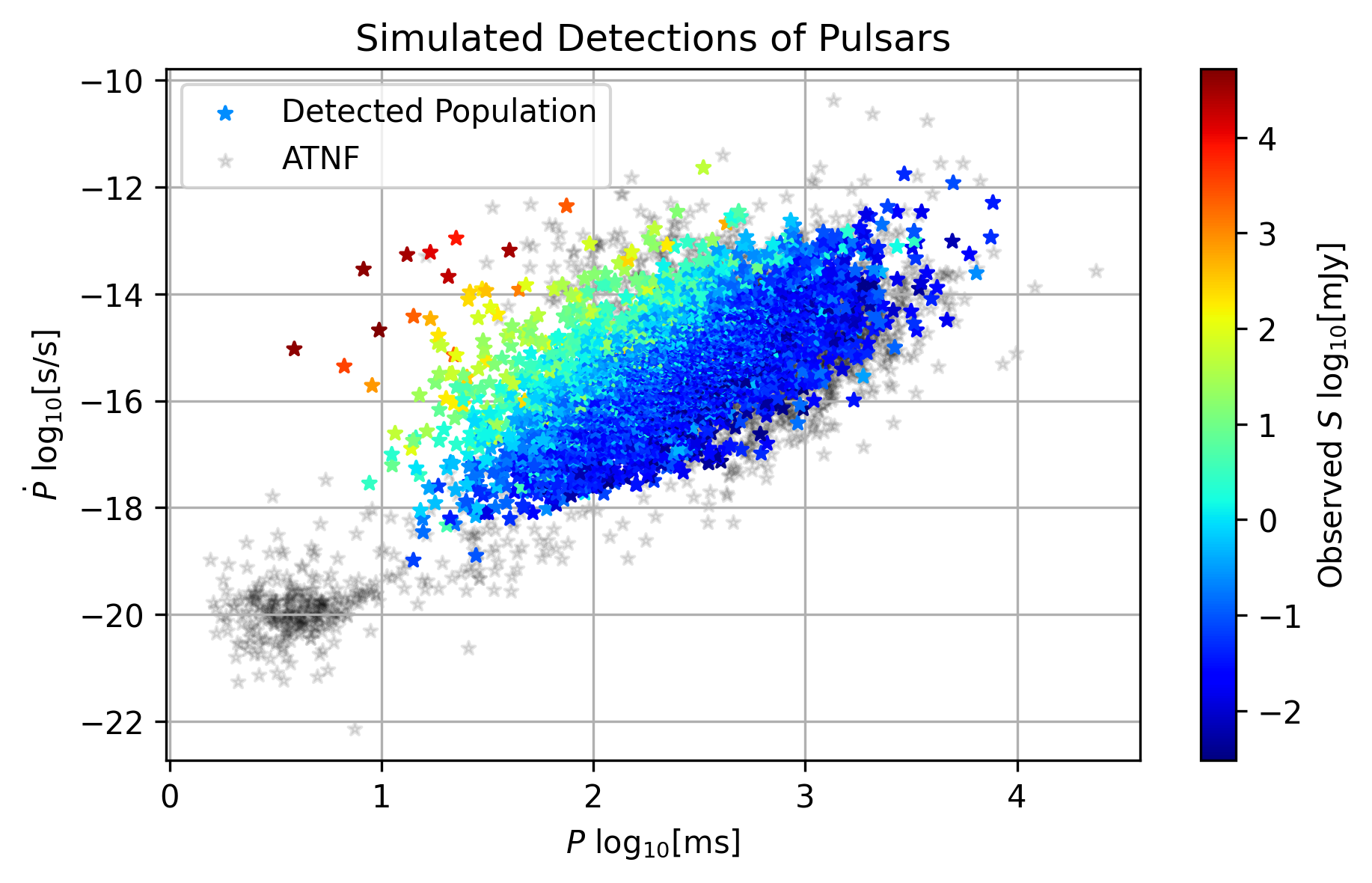}
    \caption{Assuming a beaming fraction of $15\%$ and comparing pulsar locations, fluxes, DM, and a randomized duty period, to all major surveys in the ATNF pulsar catalog results in $\sim$3,600 pulsar detections to date. Above is one such realization of detections, showing general agreement with the ATNF catalog. Note, we do not detect MSPs, as we do not model their production.}
    \label{fig:DetectedPulsars}
\end{figure}

\indent As a general sanity check, we also compare our neutron star population to all major existing pulsar surveys in the ATNF pulsar catalog \cite{ATNF_Machester_05}. We assume an average beaming fraction of $15\%$ and apply a dispersion measure, DM, proportional to distance from Earth, with some lognormal error. See Appendix \ref{app:APPLPy} for more information on this process. We find that our population produces roughly 3,600 observed pulsars to date - roughly in line with the total number of pulsars in the catalog, albeit without any MSP detections, and in line with previous population models \cite{Bhura_2024}. This is because we do not consider any spin-up processes that produce MSPs. Figure \ref{fig:DetectedPulsars} plots our detected population against the current ATNF catalog, showing general agreement in the distribution of pulsars in $P$-$\dot{P}$ space.

\subsection{A Galactic Kinematic Model}

\indent As narrow spectral lines are much easier to detect than broad lines, one needs to carefully consider Doppler-broadening of the observed background signal in the Milky Way. To this end, we consider a two-part kinematic model for neutron stars that includes the general rotation curve of the Milky Way, and the stars' kick velocities.

\begin{align}
    \vec{v}_{\text{pls}} &= \vec{v}_\text{rot} + \vec{v}_\text{kick}(t, r) \\
    |\vec{v}_\text{rot} |(0,r) &= 250 \text{ km/s} \cdot \frac{\sqrt{5 \ln (r / 2 \text{kpc} + 1)}}{r / 2 \text{kpc} + 1} \label{eq:rot_mod}
\end{align}

\indent Within a rough approximation of actual observations, we use an analytic expression for the magnitude of $\vec{v}_\text{rot}$ (Eq. \ref{eq:rot_mod}). We note that kick velocities at birth ($\approx 100$ km/s) are within an order of magnitude of or smaller than the Galactic rotation curve. To simplify orbital dynamics, we treat kick velocities as first-order perturbations that in a co-rotating frame rotate with the Galactic angular velocity, thereby generating a displacement vector from the neutron star's original orbital location. This is similar to the movement of epicyclic perturbations of thin disks. While this method does not accurately describe the motion of an individual neutron star, it is accurate in describing the ensemble motion of all stars in the Milky Way.

\begin{align}
    \vec{v}_{\text{kick}}'(t) &= \begin{bmatrix}
        \cos(\omega t) & \sin(\omega t) & 0 \\
        - \sin(\omega t) & \cos(\omega t) & 0 \\
        0 & 0 & \cos(\omega t)
    \end{bmatrix} \vec{v}_{\text{kick},0}' \\
    \vec{d}_{\text{kick}}'(t) &= \begin{bmatrix}
        \sin(\omega t) & 1-\cos(\omega t) & 0 \\
        \cos(\omega t)-1 & \sin(\omega t) & 0 \\
        0 & 0 & \sin(\omega t)
    \end{bmatrix} \frac{\vec{v}_{\text{kick},0}'}{\omega} \\
    \omega(r_0) &= |\vec{v}_\text{rot}(r_0)|/r_0
\end{align}

\indent Here, we use the apostrophe to denote a local orthogonal, spherical coordinate system where $\hat{x} \parallel\vec{v}_{rot}$, and $\hat{y} = -\hat{r}$. Notice, this simple orbital dynamics model yields a key result from more rigorous simulations and studies (e.g. \cite{Pls_Orb_Dyn}): neutron stars older than $\approx10^7$\,yr, on average, are displaced towards the outer rims of the Galaxy, as this is roughly one orbital period of the Milky Way, which is the time required for the kick displacement vector to undergo one full rotation in a co-rotating frame. There will be a small fraction of systems where $v_\text{kick}$ is much larger than $v_\text{rot}$, and where even $v_\text{pls} > v_e$. These will only affect our results by less than an order of magnitude, and one can further argue that high $v_\text{kick}$ neutron stars are partially virialized through gravitational interactions with other stars such that they no longer escape and follow trajectories more aligned with this model. Hence, we will ignore this issue for now. 

\begin{figure}[h]
    \centering
    \includegraphics[width=0.9\linewidth]{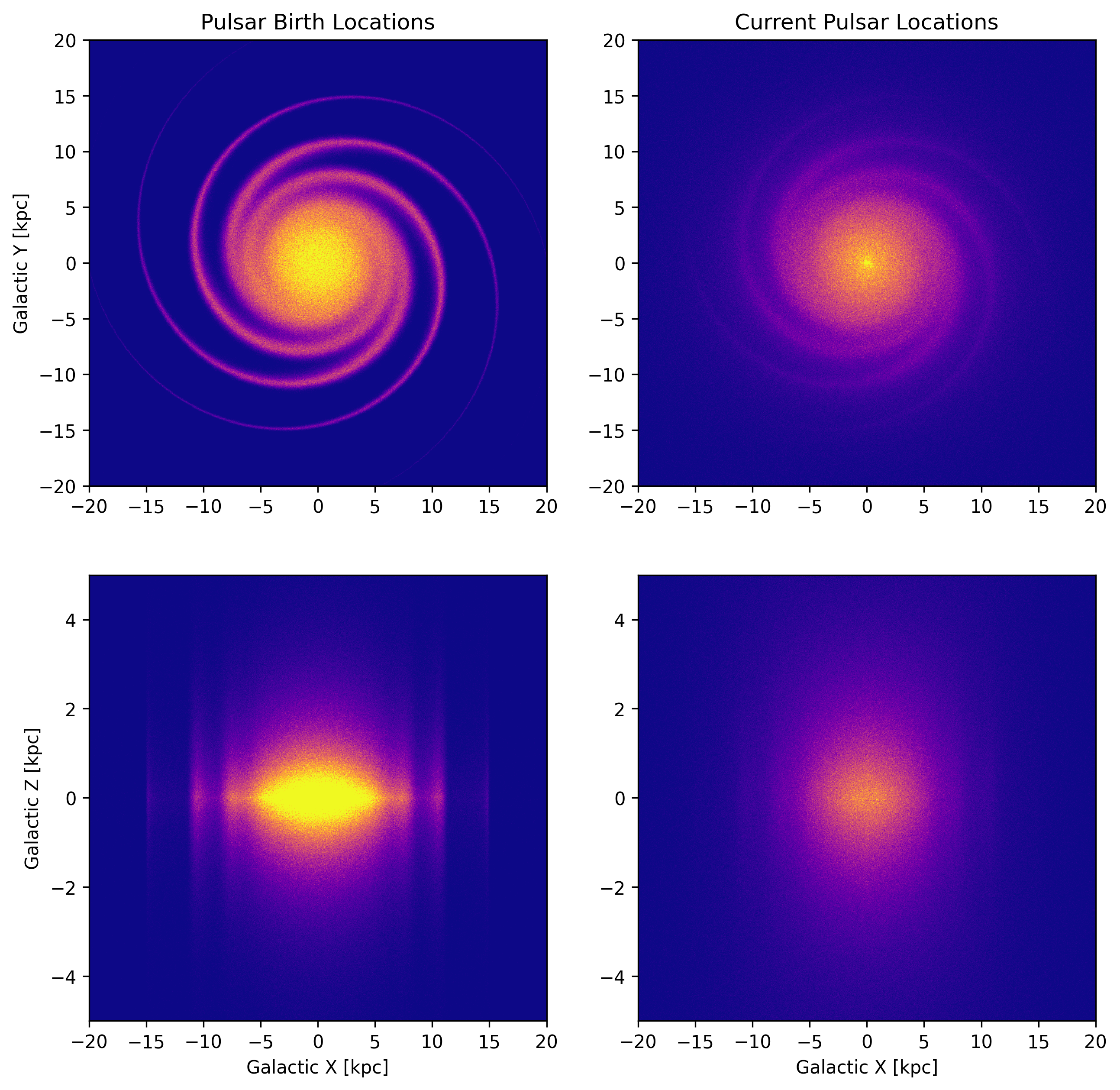}
    \caption{Left column: neutron stars histogramed by birth location. Right column: neutron stars histogramed by location at $t=0$. Colormaps are kept constant across rows to highlight the diffusion of neutron stars from star-forming regions. Note, the Z coordinate axis is exaggerated.}
    \label{fig:PulsarEvolution}
\end{figure}

\indent We choose an exponential distribution for $v_\text{kick}$ with $\mu_v = 100 $ km/s, as well as a uniformly random orientation for $\hat{v}_\text{kick}$. Computationally, it is easiest to calculate $\vec{v}_\text{kick}$ and $\vec{d}_\text{kick}$ in a local spherical coordinate system oriented with $\vec{v}_\text{rot}$ and then appropriately rotate the vectors into the coordinate system of our model. At current time, the Doppler velocity of each neutron star can then be calculated by projecting $\vec{v}_\text{pls} - \vec{v}_\odot$ in the direction of the Earth, which we assume to be $\vec{r}_\odot = \{8,0,0\}$ kpc. Figure~\ref{fig:PulsarEvolution} displays the diffusive effect neutron-star kicks have on the general population. 

\begin{figure}[h]
    \centering
    \includegraphics[width=0.99\linewidth]{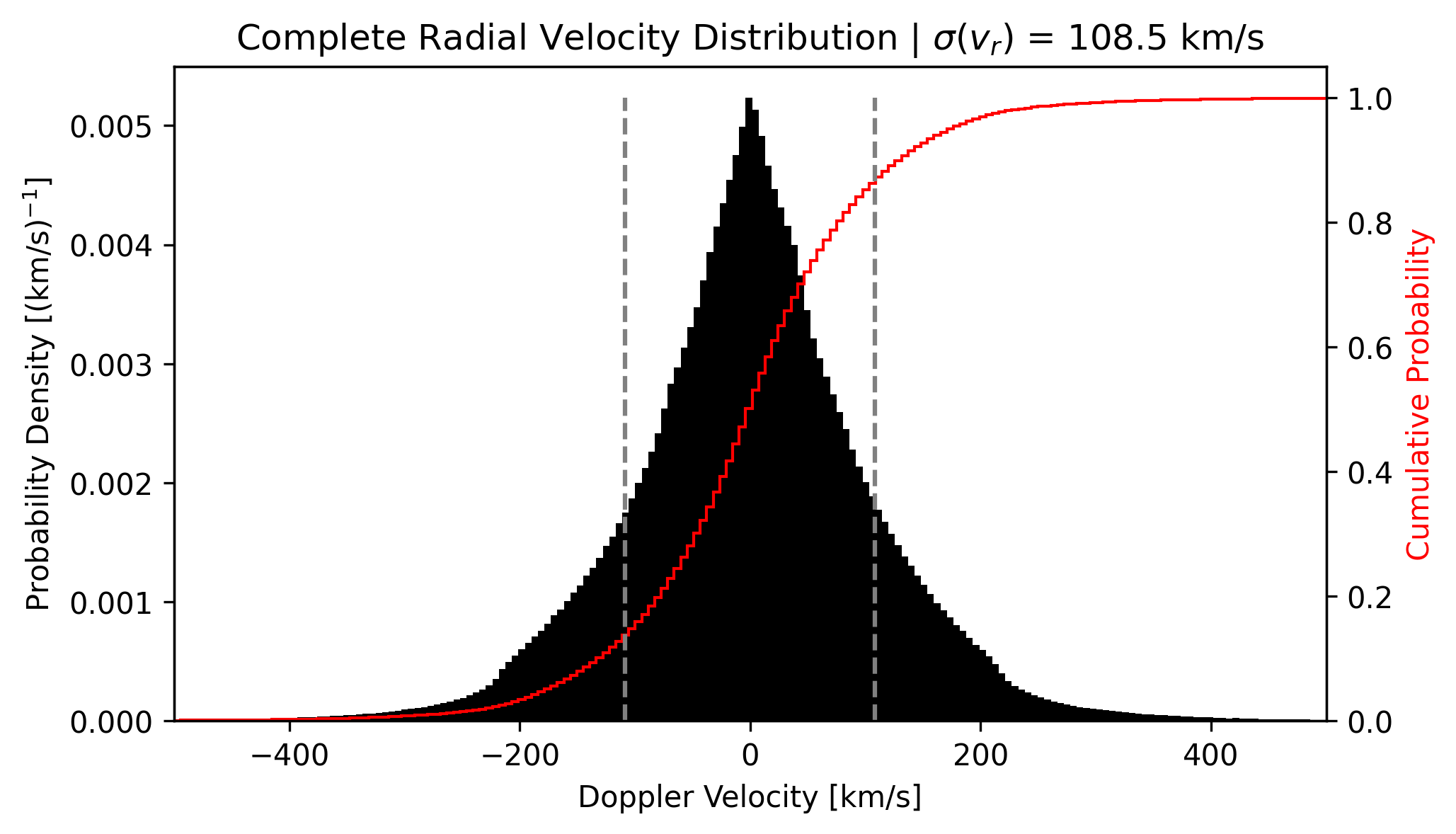}
    \caption{The complete probability density and cumulative distribution functions of the radial velocities of all neutron stars in our model. We find that the RMS radial velocity is roughly 110\,km\,s$^{-1}$ in our model.}
    \label{fig:RadialVs}
\end{figure}

\indent Figure \ref{fig:RadialVs} subsequently plots the complete distribution of radial velocities. The result is a roughly Laplacian distribution with an RMS of $\sim110$\,km\,s$^{-1}$. This implies a 3D velocity magnitude RMS of $\sim 190$\,km\,s$^{-1}$, which is well within an order of magnitude of studies like \cite{Hobbs_RadialV}.

\subsection{Calculating Background Axion Spectral Lines}

\indent One can simulate a spectral line observation by appropriately binning neutron stars by their Doppler velocities, and weighting them by their observed flux density after accounting for the antenna response patterns, which we assume to only be a function of the angular difference between a given neutron star and the orientation of the array. A sampling uncertainty for the total observed flux density in each bin can be estimated by taking an estimated standard error of the mean and multiplying by the total number of neutron stars, as we assume the fractional variance in flux is much larger than the fractional variance in the number of neutron stars sampled.

\begin{figure}[h]
    \centering
    \includegraphics[width=0.9\linewidth]{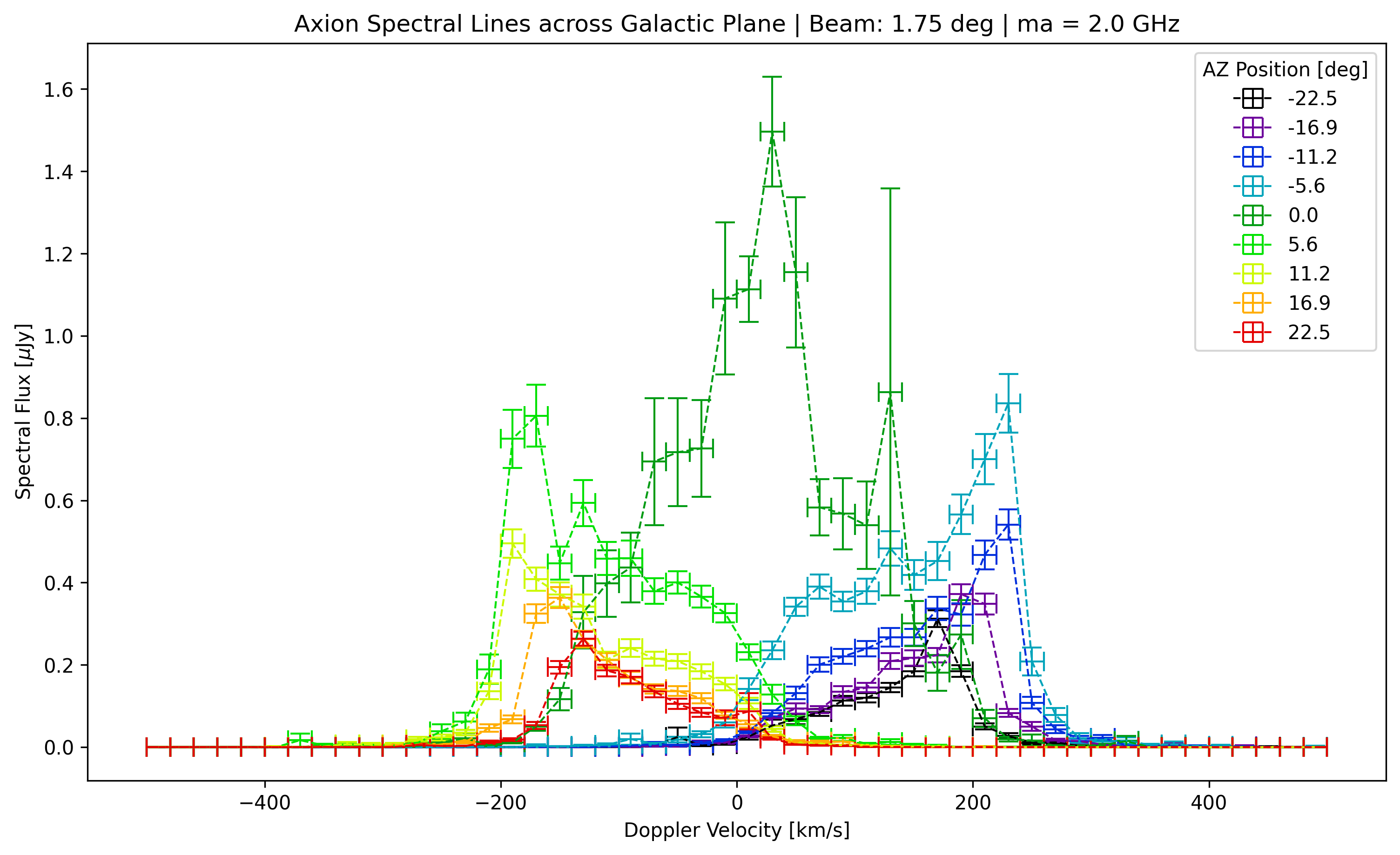}
    \caption{Simulated spectral lines from axion conversion from all neutron stars in our population model across the Galactic Center by Galactic azimuthal angle from the Earth. Total flux uncertainties are estimated from the standard error of the mean, while Doppler velocity uncertainties display the width of a given bin. In this simulation we assume $m_a/2\pi = 2$ GHz, $\Delta \nu = 20$ km/s, and 6\,m dishes. }
    \label{fig:SimAxionLine}
\end{figure}

\indent Note, the spectral flux of the theorized axion line has a $m_a^{7/3}$ scaling assuming a KSVZ model, however the antenna response pattern has an effective angular area that scales as $m_a^{-2}$, so the observed spectral flux will roughly scale as $m_a^{1/3}$ for the same instrument. Figure \ref{fig:SimAxionLine} plots simulated spectral lines across the Galactic Plane by azimuthal angle for a 2\,GHz axion mass, and an antenna response pattern determined by a 6\,m dish.

\indent Excitingly, these early results show spectral lines on the order of $\mu$Jy. However, as an effectively extended signal, these can only be detected by using autocorrelation terms of an interferometer, severely limiting the number of independent measurements available. For example, for the DSA, the thermal noise in a 1\,hr, 1\,MHz bandwidth signal from autocorrelation terms will be 680\,$\mu$Jy. This rules out direct observation of a background axion spectral line.

\indent Alternatively, one should consider the use of cross-correlation terms that are used to reconstruct radio images of the sky. On the level of individual pixels/beams in a final radio image, noise will be severely reduced as each beam will now have a thermal noise of 16\,$\mu$Jy for the same 1\,MHz spectral image. DSA survey images are expected to have upwards of 16,000,000 individual synthesized beams, allowing for sensitive estimators on the statistics of the image. Of particular interest are moment statistics, as there is a direct relationship between the distribution in beam power and a point-source background, $dN/ dS$ (See Appendix \ref{apx:cumulants}).

\subsection{Detection Techniques via Confusion and Higher-Moments of the Background}

\indent Any given astronomical image can be split between statistically significant foreground and background images via various processes. The typical delineation in power per beam between these two images is given by $5 \sigma_{\text{b}}$, where $\sigma_\text{b}$ is the standard deviation of the background image, which also constrains the uncertainty of the foreground. As a weak signal composed of many point-sources, the axion spectral line will contribute as unresolved fluctuations in the background, known as confusion noise. Unlike the axion line, typical sources of confusion are smooth across the radio spectrum \cite{Condon_2012}, as well as the radio foreground. Equation \ref{eq:imeq} is a simple model of a `cleaned' radio image that highlights this relationship, where foreground point-sources are modeled as delta functions with known spectral laws, and the background is some fluctuating field with strong correlations across radio frequency. 

\begin{equation}
    S(\vec{\theta}, \nu) = \sum_{i} S_i(\nu) \delta(\vec{\theta} - \vec{\theta}_i) + S_\text{bg}(\vec{\theta}, \nu) + S_\text{ax}(\vec{\theta}, \nu) + n_\text{th}(\nu) \label{eq:imeq}
\end{equation}

\indent Given a large-bandwidth radio image, the foreground and smooth background terms can be found and subtracted from a spectral region of interest. Best practice is to omit the small region in question when finding the foregrounds and backgrounds. This subtraction will introduce some error, however when the thermal noise in a single spectral slice is much bigger than the background confusion noise ($\sigma(n_\text{th}) = \sigma_\text{th} >> \sigma_\text{conf}$), the introduced error can typically be ignored (See Appendix \ref{app:backsub}). The resulting spectral slice will be dominated by $S_\text{ax} + n_\text{th}$.

\indent One can show that the addition of $S_\text{ax}$ will increase the cumulant statistics of the final image according to the axion background point-source spectrum $dN/dS$ (Equation \ref{eq:cummulants} - See Appendix \ref{apx:cumulants}), which can be estimated from the population model results. Specifically, the second and third cumulants are the variance and central third moments of the image, with higher moments directly related to the cumulants through Bell's polynomials. 

\begin{equation}
    \Delta \kappa_n = \Omega_\text{beam} \int S^n \frac{d S}{dN d\Omega} dS \label{eq:cummulants}
\end{equation}

\indent Staying with a thermally-limited image with $N$ beams, the uncertainty in estimates for the variance and third-central moment will be given by the following equations, which are also generalized for the $n$-th moment in Appendix \ref{apx:cumulants}:

\begin{equation}
    \begin{split}
    \sigma(\sigma^2) &= \sigma_\text{th}^2 \sqrt{\frac{2}{N}} \\
    \sigma(\mu_3) &= \sigma_\text{th}^3 \sqrt{\frac{15}{N}} \\
    \sigma_\text{th} &= \frac{2 k_B T}{A_e \sqrt{N_\text{ant}^2\Delta T \Delta \nu}} 
    \end{split}
\end{equation}

\indent The inclusion of real-world systematics, like calibration errors, side-lobe confusion and RFI noise, requires a non-trivial correction to equation \ref{eq:cummulants} and those given in \ref{apx:cumulants}. The exact details of this correction is beyond the scope of this paper, and are typically smaller than the underlying uncertainties when pixel sizes are chosen to coincide with the PSF of a given instrument (see discussion in Appendix \ref{apx:cumulants}). We therefore will continue with the example of a thermally-limited image to keep our arguments simple.

\indent Note, as the axion mass increases, the expected bandwidth of the signal increases, hence the integrating bandwidth used to detect the signal should vary with radio frequency $\nu$. Since the signal is about 300\,km\,s$^{-1}$ wide, the optimum bandwidth is on the order of $\Delta \nu \sim 10^{-3} \nu$. Secondly, $\Omega_\text{beam}$ goes as $\nu^{-2}$, so the significance (SNR) in the $n$-th cumulant statistic will have an axion mass scaling of $m_a^{17n/6 - 2}$ for a single primary beam image. An improvement can be made if one considers a constant-area survey patch, composed of multiple images. This will increase the number of beams as $\nu^2$, so the significance scaling law will be $m_a^{17n/6 - 1}$. These scaling relationships are important to keep in mind when considering multiple instruments.

\indent Survey patches are also more reliable in their statistics across radio frequency, as it is more accurate to predict $dN/dSd\Omega$ over large $d\Omega$ from our simulation, than on small patches where number counts can fluctuate strongly. We therefore consider survey images, rather than primary beam images, which are often more synergistic with other large-scale observing campaigns like those intended for DSA.

\indent One should also note an interesting consequence of the use of interferometric arrays: the scaling between $\Omega_\text{beam}$ and $A_e$ is a function of the array's baselines. Hence in a variable array, $\Omega_\text{beam}$ can be altered without changing the thermal noise per beam. Assuming $N \sim \Omega_\text{beam}^{-1}$, the SNR (which we define in units of $\sigma$: $\langle \kappa_n \rangle / \sigma(\kappa_n)$) of the cumulant estimates will then go as $\sim \Omega_\text{beam}$. This heavily favors instruments with relatively low angular resolution, but low thermal noise. On these two fronts, DSA succeeds on the later, but fails on the first. An ideal array, therefore, would have a high antennae density.

\begin{figure}[h]
    \centering
    \includegraphics[width=0.95\linewidth]{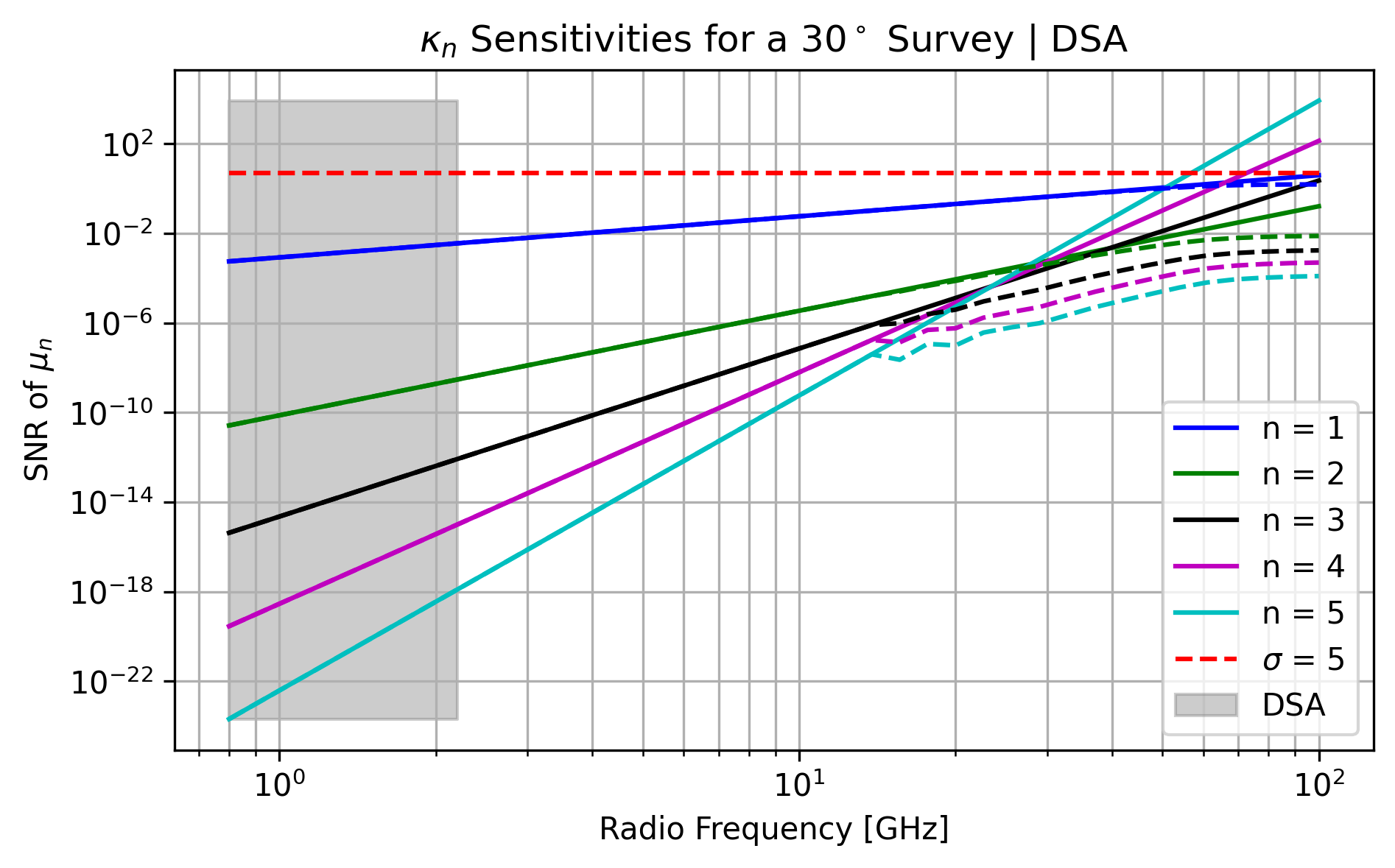}
    \caption{The expected SNR of the $n$-th cumulant statistic of the axion / neutron-star background, $\braket{\mu_n}/\sigma(\mu_s)$, over a $30^\circ \times 30^\circ$ patch centered on the Galactic Center for DSA. Telescope parameters used are: $T = 17$ K, $A_e = 9\pi \text{ m}^2$, $\Delta T = 1$ hr, $\Delta\nu = 300$ km/s, $N_\text{ant}=1650$, and $\Omega_\text{beam} = (3.3^")^2$ at 1.35 GHz. Gray denotes the operational radio range of DSA. Dotted lines show the effect of foreground removal.}
    \label{fig:DSA_Sens}
\end{figure}

\indent To illustrate the sensitivity of different instruments to a KSVZ axion using these higher order methods, we plot the SNR of the expected increase in each moment statistic of the background for DSA (Fig. \ref{fig:DSA_Sens}). Given survey images integrated over one hour, it is clear that typical SNRs within the DSA's operational radio range are far too low ($< 10^{-5}$) for any reasonable detection of the axion background using the instrument. On the other hand, it is clear that SNRs increase drastically with radio frequency, highlighting a need for an array at submillimeter wavelengths.

\begin{figure}[h]
    \centering
    \includegraphics[width=0.95\linewidth]{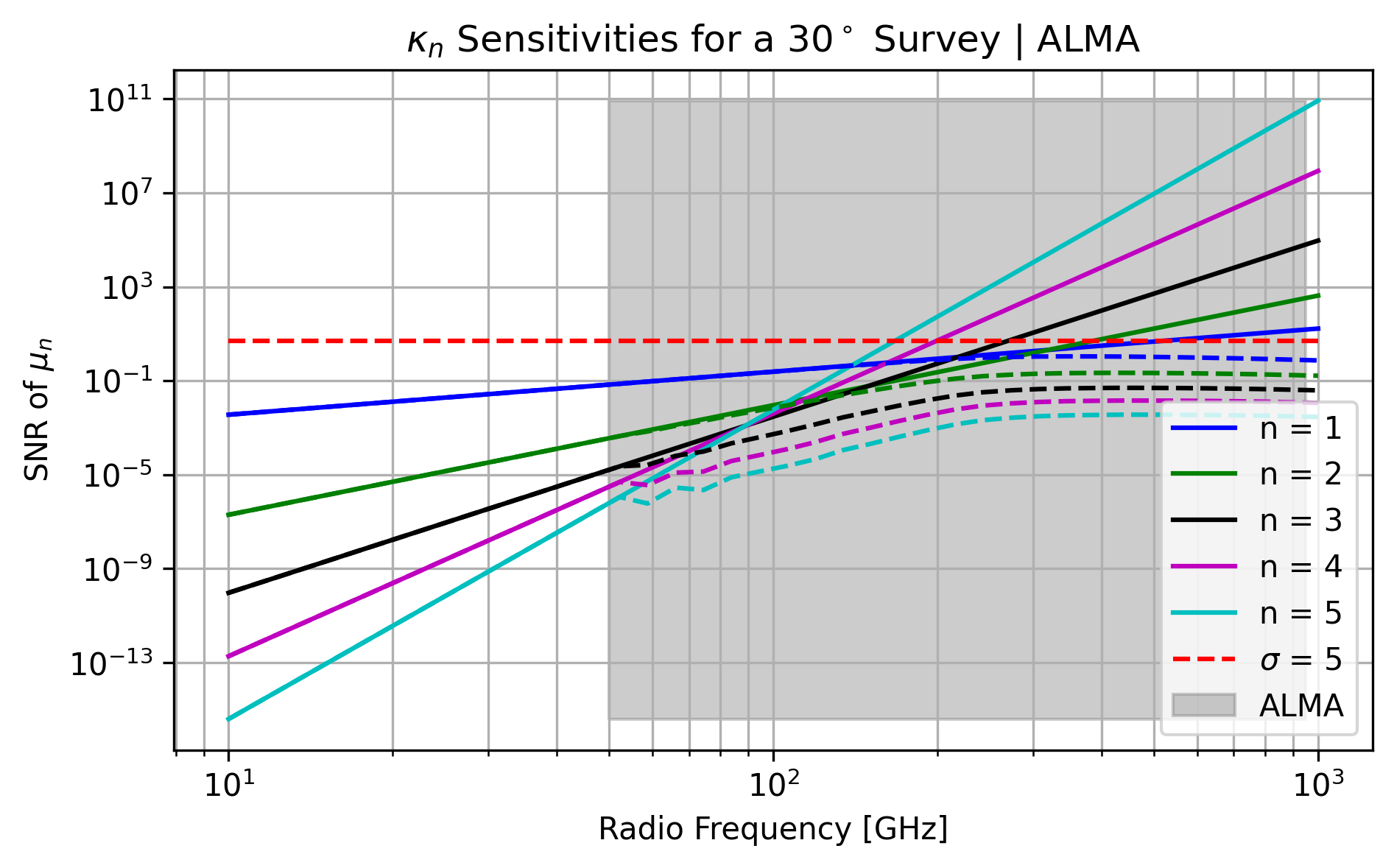}
    \caption{The expected SNR of the $n$-th cumulant statistic of the axion / neutron-star background, $\braket{\mu_n}/\sigma(\mu_s)$, over a $30^\circ \times 30^\circ$ patch centered on the Galactic Center for ALMA. Telescope parameters used are: $T = 100$\,K, $A_e = 36\pi \text{ m}^2$, $\Delta T = 1$\,hr per field, $\Delta\nu = 300$\,km\,s$^{-1}$, $N_\text{ant}=66$, and $\Omega_\text{beam} = (0.5^")^2$ at 950\,GHz. Gray denotes the operational radio range of ALMA. Dotted lines show the effect of foreground removal.}
    \label{fig:ALMA_Sens}
\end{figure}

\indent To this end, one can perform similar calculations for the 66-element ALMA, which operates from 35 GHz to 950 GHz, and has variable baselines. In the most compact configuration, the telescope has a resolution of 0.5$^"$ at 950 GHz – effectively lower resolution than DSA at comparable radio frequencies. Figure \ref{fig:ALMA_Sens} shows that at higher radio frequencies, the axion / neutron-star background would be detectable with ALMA. In particular, higher-moment statistics are more sensitive at these higher scales. We note that our model does show that some neutron stars will be detectable in the foreground at these radio frequencies; the dotted curves in Fig. \ref{fig:ALMA_Sens} show the effect of their removal as part of the foreground. However, searching for spectrally-narrow point sources is a computationally and observationally intensive task, with high false-alarm probabilities. The use of higher-moment statistics nicely complements such a search, as they will highlight at which radio frequencies one should search for individual neutron stars, while constraining false-alarm probabilities. 

\indent In this paper, we have largely considered the KSVZ axion, with a strict relationship between $m_a$ and $g_{\alpha \gamma}$. For completeness, we will let $g_{\alpha \gamma}$ vary to highlight which sectors of the axion-photon coupling different instruments can probe. Under the assumption of a m-$\sigma$ detection threshold, a sensitive coupling constant can be computed from the previous SNR results under different moment statistics (Eq. \ref{eq:g_sens}), ignoring any foreground subtraction. 

\begin{equation}
    g^{(n)}_{\alpha \gamma, \text{sens}} = \left[ \frac{m_a^2}{\text{SNR(n)}^2}\right]^{1/4n} g_{\alpha \gamma, \text{KSVZ}}(\nu) \label{eq:g_sens}
\end{equation}

\begin{figure}
    \centering
    \includegraphics[width=0.95\linewidth]{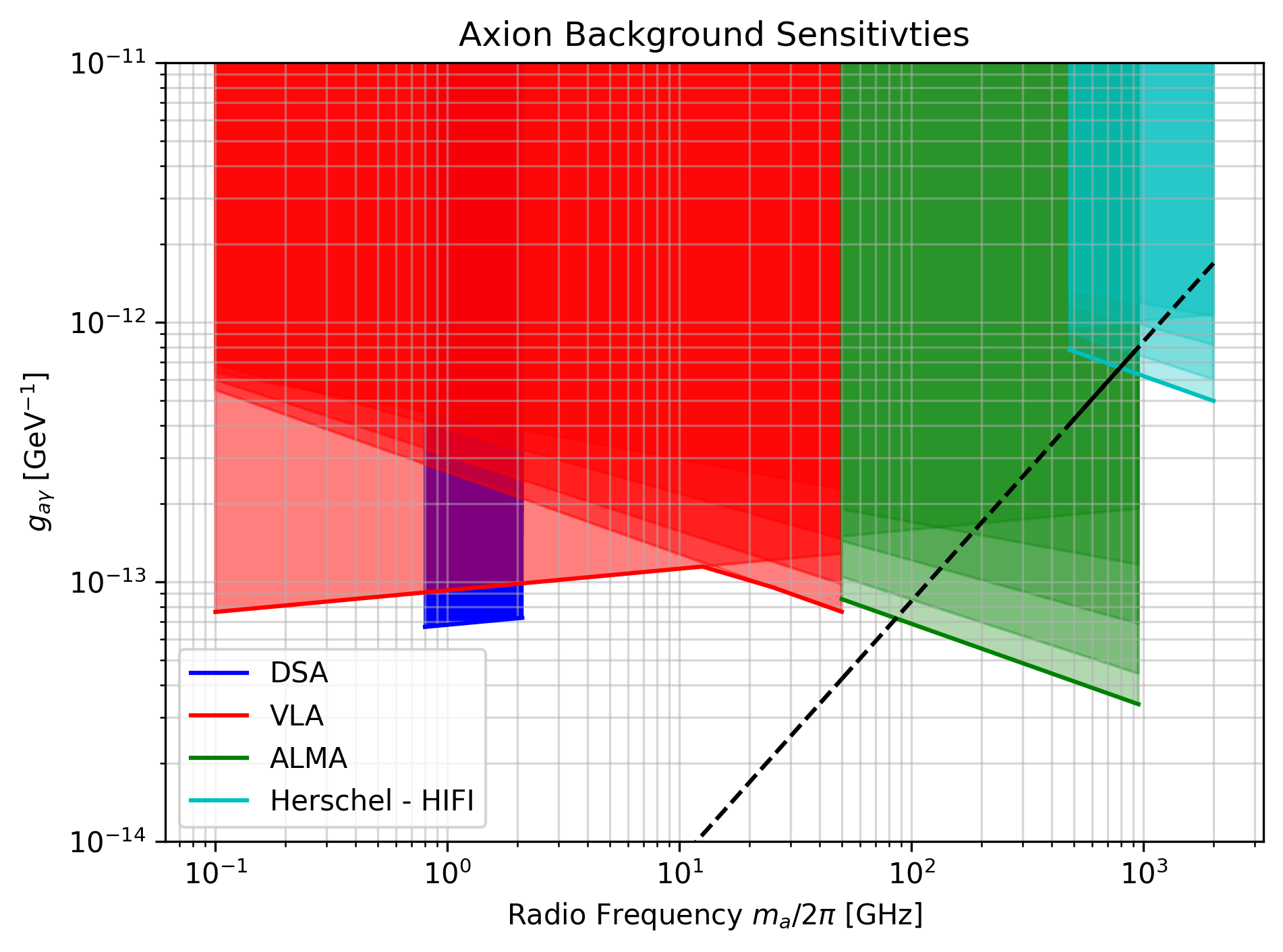}
    \caption{5-$\sigma$-sensitive sectors of the axion coupling constant $g_{\alpha \gamma}$ for different instruments under the same $30^\circ \times 30^\circ$ survey parameters for a thermally limited image. Here we assume a primary-beam image integration time of 1 hour, and a bandwidth integration resolution of 300\,km\,s$^{-1}$. See Appendix \ref{apx:instinfo} for individual instrument parameters used. The dotted black line denotes the KSVZ axion.}
    \label{fig:AxionSens}
\end{figure}

\indent Figure \ref{fig:AxionSens} plots the resulting sensitivity curves for different instruments in the radio regime under a 5-$\sigma$ threshold. Excitingly, ALMA would be able to probe current theories of the axion in the sub-millimeter regime using higher order methods, pushing beyond previously computed sensitivities at lower radio frequencies, such as in \cite{bhura2026axionsearchtelescoperadio}. Interestingly, the three modern radio instruments considered here - DSA, VLA, ALMA – tend to be similarly sensitive to a coupling constant between $10^{-13}$ and $10^{-12}$ GeV$^{-1}$ across their respective radio frequency windows. However, this disadvantages instruments that operate at lower radio frequencies in the search for axions. For this reason, we also investigate the use of an instrument like the Herschel Space Observatory, which operates in the near-IR. While being an extremely cold instrument ($\le 3$\,K), it is limited in it's inability to conduct interferometric measurements that greatly reduce the effect of thermal noise per beam, however for similar large-scale surveys, this makes it capable of detecting background statistics.

\begin{figure}[h]
    \centering
    \includegraphics[width=0.95\linewidth]{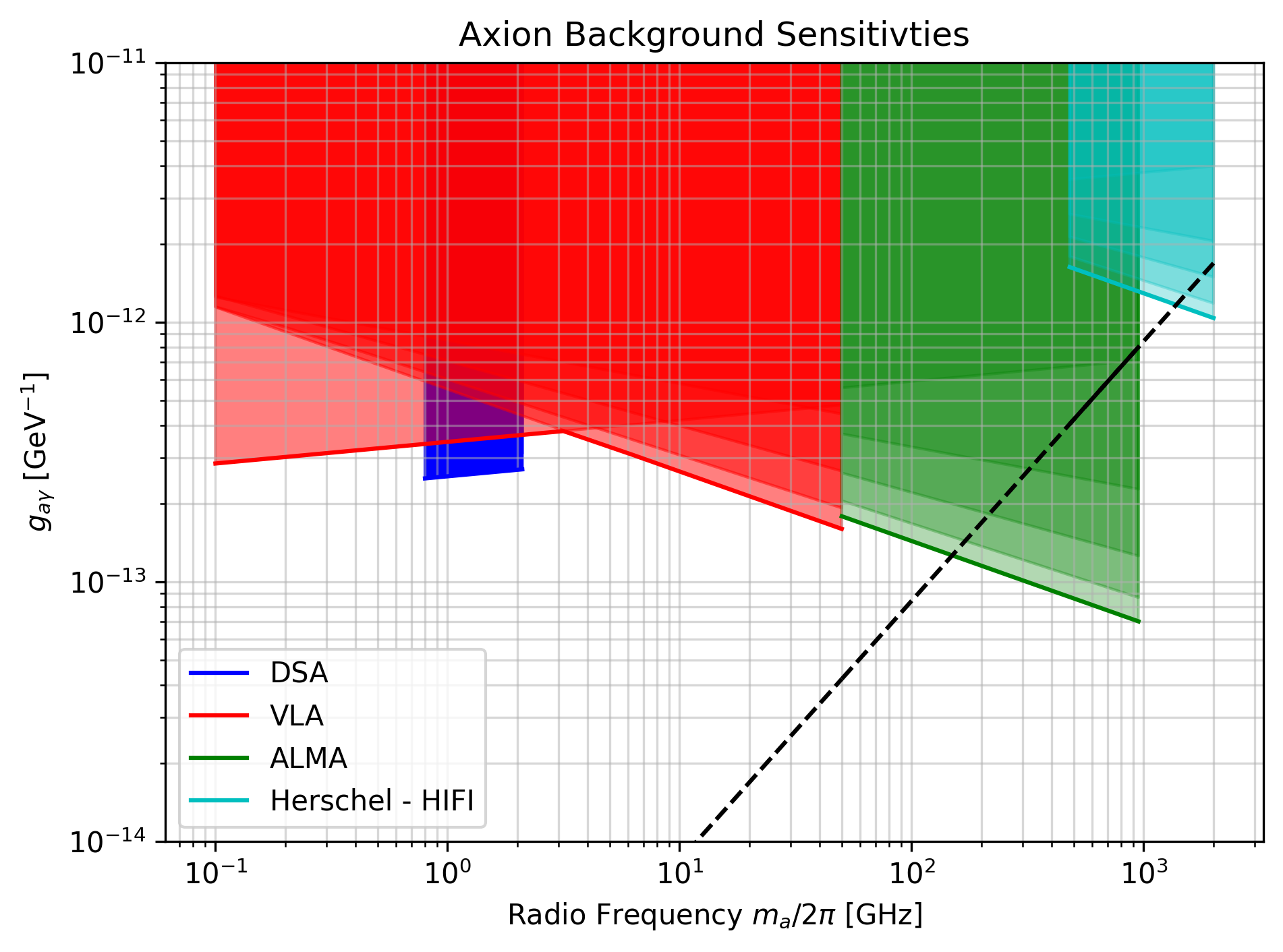}
    \caption{5-$\sigma$-sensitive sectors of the axion coupling constant $g_{\alpha \gamma}$ for different instruments under a more realistic $1.5^\circ \times 1.5^\circ$ survey with integration times of 1\,min.}
    \label{fig:AxionSens_Realistic}
\end{figure}

\indent We will note that the survey parameters used in Fig.~\ref{fig:AxionSens} do not make practical sense for a non-survey instrument like ALMA or Herschel. The total time to observe a $30^\circ \times 30^\circ$ patch for beam integration times of $1$\,hr will be on the order of $10^6$\,hr for ALMA – an impractical time frame for any campaign. By comparison, such a survey would only take $\sim$500\,hr with DSA. By the same estimates, one can show that a $1.5^\circ \times 1.5^\circ$ survey with $\Delta T = 1$\,min will take only 100\,hr. Figure \ref{fig:AxionSens_Realistic} shows a significant reduction in sensitivity for lower moment statistics, with total power measurements no longer sensitive to the axion background. Confusion and higher moment estimates, however, will still be sensitive to the axion background at high radio frequencies ($>300$\,GHz) for ALMA, highlighting their usefulness in quickly resolving effects of the background in any realistic observing campaign. The same reduced survey parameters for Herschel, however, still require a large 100 days to complete.

\indent We also estimate our systemic uncertainties for our $g_{\alpha \gamma}$ sensitivity curves between 0.9 to 2.2 dB across EM frequency and instrumentation, but dependent on the cumulant statistic method used. Table \ref{tab:ResultsUncerts} gives these logarithmic uncertainties for $g_{\alpha \gamma}$ and SNR estimates from 100 realizations of our population model. Note that uncertainties for SNR estimates are much higher for higher-order cumulants, but are much more sensitive to changes in $g_{\alpha \gamma}$. Hence uncertainties in the coupling sensitivity levels are much smaller.

\begin{table}[h]
    \centering
    \begin{tabular}{|c||c|c|c|c|c|} \hline
        $\sigma$ [dB] & $\kappa_1$ & $\kappa_2$ & $\kappa_3$ & $\kappa_4$ & $\kappa_5$ \\ \hline \hline
        $g_{\alpha \gamma, \text{sens}}$ & 0.93 & 1.84 & 2.05 & 2.12 & 2.15 \\ \hline
        SNR & 1.85 & 7.37 & 12.28 & 16.94 & 21.50 \\ \hline
    \end{tabular}
    \caption{Estimated systemic uncertainties in $g_{\alpha \gamma}$ sensitivity levels and SNRs for each cummulant method from 100 model realizations. Uncertainties are computed and quoted in dB, and are instrument and $m_a$ independant.}
    \label{tab:ResultsUncerts}
\end{table}

\subsection{Individual Point Sources $\&$ Variance in Statistical Predictions \label{sec:surveymethod}}

\indent We have largely ignored the resolution of individual point sources for three reasons: the possibility of doing so has been explored in \cite{dsa_std_mod} in the DSA radio regime; due to the large uncertainty in the underlying assumptions of our neutron star population model, strict predictions are hard to constrain; in large-scale surveys, false-alarm probabilities significantly hamper the possible detection of individual dim sources. The general consensus in \cite{dsa_std_mod}, in agreement with ours, is that if an axion theory were valid, there is a good chance that there exists a pulsar in our galaxy with a 5$\sigma$-significant axion line in a thermally-limited image. In fact, in the `realistic' survey considered in Fig. \ref{fig:AxionSens_Realistic}, we find a steady increase in individually observable axion lines starting at 150\,GHz, up to $10^3$ observable point sources at 950\,GHz for ALMA.
That being said, this very same survey with ALMA, one expects $\sim10^1-10^2$ beams to exceed the $5\sigma$ threshold \emph{per} integrated bandwidth image – any slight increase in unmodeled noise, especially skewed noise, will quickly increase this number by orders of magnitude. This puts a significant limit on the efficacy on an individual point-source search, requiring an extremely bright point source to overcome the false-alarm probability thresholds of any complete survey. The only way to constrain false-alarm probabilities otherwise would be through specialized follow-up observations.

\indent The probability of the existence of such a source we leave as an open problem, and presents an interesting application of statistics. On the other hand, we do note that these same uncertainties in the distribution of different pulsar parameters – even the total number of pulsars in the Milky Way – affect our own results. The heuristic model we have developed assumes $\sim 10^7$ total neutron stars, alongside specific prior distributions for $B_0, P_0, v_\text{kick}$ and their birth locations. Some have estimated the actual number of neutron stars in the Milky Way as high as $10^9$, which increases the expected strength of different background moment estimators by up to a factor $10^2$. On the other hand, assumptions in $B_0$ and the $90^\circ$ inclination angle result in a high number of neutron stars with large $B_0$ and $P$, which may bias our moment estimates high. However, as a whole, we believe our results to be heuristic to what one expects for the Milky Way.

\indent We therefore believe that the most effective methodology in searching for the axion spectral line will combine the higher-order methods outlined in this paper alongside a standard point-source search as suggested in \cite{dsa_std_mod}. First, a general survey should be conducted to calculate deviations in background moments across radio frequency. The bandwidths that display the highest deviations should then be searched for potential point sources, and/or followed up with a more sensitive observing campaign in the search for individual sources and to better constrain the statistics of the background.
  
\section{Axion-Charged Particle Interactions in the ISM \label{sec:axioninism}}

\indent We now explore axion conversion in interactions with the individual electromagnetic fields of free charged particles. First, we will need to calculate all the relevant cross sections by considering a direct axion-particle (Primakoff) process, as outlined in \cite{Wu_2024}, with a matrix element given by

\begin{equation}
    |\mathcal{M}|^2 = g^2|\epsilon_\mu\varepsilon^{\mu \nu \alpha \beta}  p_{\gamma,\alpha} q_\beta \frac{e}{q^2} F(q) [\bar{u}_i \gamma_\nu u_f]|^2,
\end{equation}

\indent where $p_\gamma$ is the outgoing photon 4-momentum, $q$ is the exchanged 4-momentum, $F(q)$ is the relevant form factor of the particle's electric field, $u_{i/f}$ are the initial and final Fermion states, $\epsilon$ is the photon polarization vector, and $g$ is the $g_{\alpha \gamma \gamma}$ coupling constant. In the low-energy regime, one normally takes $F(q) \approx 1$, however we introduce a correction for ionized plasmas, as free particles will be screened according to the local Debye length:

\begin{equation}
    F(q) = \frac{|q|^2}{|q|^2 + \kappa_s^2}, \text{ } \kappa_s^2 = \left[\frac{\hbar^3}{c^5} \right]\frac{4 \pi \alpha}{k_BT} \left(\sum_i Q_i^2 n_i\right).
\end{equation}

\indent Note, in this section we will assume natural units. In the C.O.M. frame, the differential cross section can be computed using the C.O.M. mass squared $s$:

\begin{equation}
    \begin{split}
    \frac{d\sigma}{d\Omega} = &\frac{g^2}{64 \pi^2 s} \frac{\alpha}{ q^2} \left( \frac{|q|^2}{|q|^2 + \kappa_s^2(T,n_i)}\right)^2  \times \\ 
    & \times\left|\epsilon_\mu\varepsilon^{\mu \nu \alpha \beta}  p_{\gamma,\alpha} q_\beta [\bar{u}_i \gamma_\nu u_f]\right|^2 
    \end{split} \label{eq:diffcrossdec}
\end{equation}

\indent Note, assuming that the axion mass is much less than the interacting charged fermion mass will result in a photon with energy $m_a \sqrt{1 - \Delta v^2}$ in the C.O.M. frame, where $\Delta v$ is the relative velocity difference between the interacting particles.  

\indent The final step in computing the cross section of this process is to average over each photon-spin and charged-particle-spin state combination, and evaluate for $g$. In this paper we will assume for $g$ that $E/N = 0$, as aligned with the KSVZ model. This assumption gives the largest values for $g$ within current QCD Axion theories \cite{di_Cortona_2016}. Generally, one should also allow this to vary, but for heuristic purposes, we make this simplification.

\indent In the low-energy regime, one will find that the differential cross section is dipolar with equal amplitudes for forward and backward scattering of the axion into a photon. We therefore apply a numerical integrator to compute the total cross section at differing COM momenta using Eq.~\ref{eq:diffcrossdec}. At highly relativistic momentum scales, care needs to be taken in this calculation as the differential cross section will approach a delta function around $\theta = 0$. We find that as long as the C.O.M. axion momentum is less than the axion mass, the total cross section is independent of any velocity terms. Using $T= 8000$\,K and $n_e = 1$\,cm$^{-3}$ \cite{Driane}, we find the following approximate relationship in the non-relativistic regime for this typical plasma scale:

\begin{align}
    \sigma \approx 4.56 \times10^{-52} \text{ cm}^2 \cdot \left( \frac{m_a}{\text{eV}}\right)^2 \label{eq:crossapprox}
\end{align}

\indent When $p_a/m_a < 5\%$ in the COM frame, this cross-section is valid to within $<1\%$, and therefore a good approximation in the non-relativistic regime. Note, when approaching relativistic regimes - up to $p_a / m_a < 50\%$ - a post-Newtonian correction factor of $\sim [1 + 2(p_a/m_a)^2]$ can be applied to ensure the same $<1\%$ agreement. Any further relativistic correction terms are outside the context of this paper.

\indent For a classical plasma, another important correction term is Debye screening, which will be a function of $n_e/T$, and is strongest for small $q^\mu q_\mu$ - i.e., forward scattering. This adds a non-trivial dampening factor to the total cross-section that can be integrated out. At a heuristic level, we therefore expect screening to only become significant when the Debye length is of similar scale to the least momentum transfer: $\kappa_s^2 \sim \min(q_\mu q^\mu)$. In the non-relativistic regime, $|q^2|$ will be given by $2m_a^2$ and can be treated as independent of the scattering angle $\theta$. Eq.~\ref{eq:debyefactor} then gives the resulting correction for Debye screening. 

\begin{equation}
    \sigma(n_e,T,m_a) \approx \sigma(m_a) \cdot \left[1 + \frac{n_e}{T}\frac{4 \pi \alpha e^2}{m_a^2} \right]^{-1} \label{eq:debyefactor}
\end{equation}

\indent One can therefore conclude that Debye screening only becomes significant when $n_e/T \gtrsim m_a^2 / (4\pi \alpha e^2)$. For a $10^{-6}$\,eV axion (corresponding to the radio range), this limit is given by 0.12\,cm$^{-3}$\,K$^{-1}$. Hence, in most astrophysical plasmas, especially in diffuse plasmas, Debye screening will not be significant. Therefore the approximation given by Eq.~\ref{eq:crossapprox} is effectively constant over a wide order of magnitudes in $T$ and $n_e$ and still correctly approximates cross sections in non-relativistic interactions at baryon number densities well beyond the redshift at recombination ($z > 1000$). We therefore adopt this cross-section for the rest of this paper, and apply the Debye correction factor when relevant.

\subsection{The Axion Spectral Line in a Classical Plasma}

\indent While the cross-section of the considered process is extremely small, one may expect this to be counteracted by a system of some considerable size like the interstellar medium. To first-order, the total photon production rate from this process will be $\Gamma \approx \sigma n_e n_a \langle\Delta v \rangle$. Reformulating this in terms of a dark matter density, $\rho_\text{DM}$, and a luminosity density gives the total energy output of this process:

\begin{equation}
    \begin{split}
    \frac{d\dot{E}}{dV} \approx 0.857 &\text{ W  pc}^{-3}\cdot \left( \frac{m_a}{\text{eV}}\right)^2  \left( \frac{n_e}{\text{cm}^{-3}}\right) \times \\
    &\times\left( \frac{\Delta v}{10 \text{ km/s}}\right) \left( \frac{\rho_{DM}}{0.4 \text{GeV / cm}^3}\right)
    \end{split}
\end{equation}

\indent However, the more useful radiometric quantity is the spectral flux density that these processes could produce. This is largely dependent on the exact geometry of the momentum space of the axion and fermions; however, we expect luminosity to be at a minimum when there is no \emph{net} flow between the two species. We therefore make the ansatz of a thermally distributed plasma and a cold, low-energy axion distribution, with axions distributed around the origin by a delta function, and charged particles following a Maxwell-Boltzmann distribution. This leads to a `nice' geometry where the C.O.M. momentum of the particles, and the rest-frame total momentum are parallel. Applying a dipole approximation to the differential cross section, $d\sigma / d\Omega \approx 3 \sigma /2 \cdot \cos^2(\theta)$, one can find the resulting spectral differential cross section for the observed energy of the emitted photon:

\begin{equation}
    \begin{split}
    \frac{d\sigma}{dE} = \frac{3 \sigma}{2 \gamma v^3 m_a}&\left(1 - \frac{E}{\gamma m_a} \right)^2
    \approx \frac{3\sigma(m_a - E)^2}{2 m_a^3 v^3}
    \\
    \\ &\text{for } \text{ }  |E/m_a - 1| \leq v
    \end{split}
\end{equation}

\indent The quadratic (non-relativistic) approximation above can then be integrated against the Maxwell-Boltzmann distribution of $v$, giving the resulting spectral photon rates (taking $m_i$ to me the mass of the interacting charged particle):

\begin{equation}
    \begin{split}
    \frac{d\Gamma}{dE} = \frac{3 n_i n_a \sigma}{2 m_a}&\left( \frac{E}{m_a} - 1\right)^2 \left[ \frac{m_i c^3}{k_B T}\right] \times\\
    &\times\text{ Erfc}\left( \left| \frac{E}{m_a} - 1\right| \sqrt{\frac{m_i c^2}{k_B T}}\right)
    \end{split}
\end{equation}

\indent This distribution has a doublet structure because forward and backscattering is possible in the conversion process. The peak $\Gamma$ and photon energy can then be calculated directly using $\xi_0 = 0.841882:$

\begin{align}
    E_{\text{peak}} &= m_a \left[ 1 \pm \xi_0 \sqrt{\frac{k_B T}{m_i c^2}} \right]  \\
    \frac{d\Gamma_\text{peak}}{dE} &= \frac{3c n_i n_a \sigma}{2 m_a} \cdot \xi_0^2 \\ 
    \frac{dL_\text{peak}}{df dV} &= \frac{3hc}{2} n_i n_a \sigma \cdot \xi_0^2
\end{align}

\indent Note that the peak spectral rates and luminosities are independent of the thermal properties of the plasma, however the Fermion mass and temperature will shift the location of the observed peaks. For example, for an ionized hydrogen plasma, we would effectively see three peaks: two peaks either side of the axion mass frequency from interactions with free electrons, and a central peak (a very narrow doublet) from interactions with the free protons. 

\indent To check if these rates are detectable, we heavily approximate the Milky Way ISM around Earth as a plasma sphere of radius 1\,kpc, with $n_e \approx 1$\,cm$^{-3}$. The local spectral radiance will then be:

\begin{equation}
    \begin{split}
    L_{f,\Omega,\text{peak}} = \frac{1 \text{ kpc}}{4 \pi}& \frac{dL_\text{peak}}{df dV} = 9.44 \times 10^{-16} \text{ Jy/str} \times  \\
    &\times\left( \frac{m_a}{\text{eV}}\right)\left( \frac{\rho_{DM}}{0.4 \text{GeV / cm}^3}\right)
    \end{split}
\end{equation}

\indent Considering the current upper limit of $m_a$ at \emph{roughly} $10^{-2}$\,eV\,c$^{-2}$ \cite{AxionUpperBound}, this rules out an observation of this process in the local Milky Way from the ISM. Even if one considers changes to the axion-Fermion geometry considered in this example, such as the ISM co-rotating with the Milky Way through the dark matter halo, we expect these changes to alter spectral radiance by at most one to two orders of magnitude. 

\subsection{Axion-Photon Conversion in the Milky Way and Local Universe}

\begin{table*}
\begin{ruledtabular}
\begin{tabular}{|c||c|c|c|c||c|}
        Object & $\langle n_e \rangle$ [cm$^{-3}$] & $\langle n_a \rangle$ [cm$^{-3}$] & $\langle \sigma \rangle$ [cm$^{2}$] &$\Delta r$ & $L_{f, \Omega, \text{peak}}$ [Jy/str] \\  \hline 
        Solar Photosphere \cite{SolarBook} & $10^{17}$& $3\times10^{13}$& $10^{-72}$& 400 km & $2\times10^{-27}$\\
        Sgr B2 \cite{SgrB2} &$3\times10^3$ & $3\times10^{15}$ & $5\times 10^{-62}$& 50 pc& $2\times10^{-17}$\\ 
        Sgr B2 at Sgr A$^*$ & $3\times10^3$&$10^{17}$ &$5\times 10^{-62}$ & 50 pc & $7\times10^{-16}$ \\
    \end{tabular}
    \caption{First-order peak spectral radiances from axion-photon conversion in known, ionized astrophysical plasmas assuming a $10^{-5} $ eV axion.\label{tab:PlasmaRad}}
\end{ruledtabular}
\end{table*}

\indent Relative to many other astrophysical objects, the ISM has a low electron number density, and our local section of the Milky Way has a low dark matter density when assuming standard DM halo models. We therefore now explore the resulting brightness temperatures of different astrophysical contexts. 

\indent If we assume a certain region is roughly isothermal, the resulting peak spectral radiance seen by an observer can be calculated as an integral along the line of sight and also further approximated over a limited area:

\begin{equation}
    \begin{split}
    L_{f, \Omega, \text{peak}} &= \frac{3 h c \xi_0^2}{8 \pi}\int n_e n_a \sigma dr  \\
    &\approx \frac{3 h c \xi_0^2}{8 \pi} \braket{n_e} \braket{n_a} \braket{\sigma} \Delta r
    \end{split}
\end{equation}

\indent As an example, if we consider the solar photosphere, up to a depth of 400\,km, and a free electron density of $10^{17}$\,cm$^{-3}$, the resulting spectral radiance \footnote{Note: within these approximations, spectral radiance will follow a $m_a^2$ proportional scaling.} for a $10^{-5} $\,eV axion will be $10^{-17}$\,Jy/sr using the standard cross section found earlier. However, in the case of the Sun, the true cross section is significantly reduced due to Debye shielding, and the spectral radiance drops by over 10 orders of magnitude. This highlights a need for higher (but not too high) densities at large scales. Table \ref{tab:PlasmaRad} gives some examples of heuristic astrophysical objects most likely to produce a detectable signal.

\indent The axion number densities in Table \ref{tab:PlasmaRad} come from roughly applying a NFW model \cite{Navarro_1996}, however they are only meant to be accurate to an order of magnitude. The general result is that even for the most extreme objects, any expected spectral lines will be too weak to detect, with current radio technology requiring a signal of at least 1\,$\mu$Jy for a significant detection. For example, the Central Molecular Zone has an increased dark matter density that is $10^2 - 10^3$ times higher than at the Earth. For Sgr B2 specifically, assuming full ionization only brings spectral radiance up to $\approx10^{-17}$\,Jy/sr. Even if one 'places' the cloud right at Sgr A$^*$ and integrates a line of sight path close to Sgr A$^*$, the spectral radiance increases only to $\sim10^{-15}$\,Jy/sr.

\indent At cosmological scales we will now note a useful upper limit of the spectral radiance over any line of sight, assuming $\sigma$ is constant over the path in consideration. By the Cauchy-Schwarz inequality, $L_{f,\Omega,\text{peak}}$ is bounded above by the variance in matter density fluctuations:

\begin{equation}
    L_{f, \Omega, \text{peak}} \leq \frac{3 h c \xi_0^2 \sigma}{8 \pi} \braket{n_e}\braket{n_a}\Delta r \left[( \braket{\delta_e^2} + 1)( \braket{\delta_a^2} + 1)\right]^{1/2}
\end{equation}

\indent This is entirely general, as long as $\sigma$ remains constant. We can therefore extend our line of sight out to near cosmological scales ($\approx 10$\,Mpc) such that we can mostly ignore the effect of redshift while still making use of large-scale statistics of matter density fluctuations $\delta_i$. If we assume baryon densities and free electron densities to be directly correlated, we can write this limit using $\Lambda$CDM parameters, where $\mu_e$ is the mean molecular weight per free electron, in the limit of large fluctuations:

\begin{align}
    L_{f, \Omega, \text{peak}} \leq  \frac{3 h c \xi_0^2 \sigma}{8 \pi}  \frac{\Omega_b \Omega_c}{\mu_e m_a} \rho_c^2 \Delta r \cdot \sqrt{ \braket{\delta_b^2} \braket{\delta_a^2}}
\end{align}

\indent If one works through the pre-factor in the above equation, one quickly realizes that for $L_{f,\Omega,\text{peak}}$ to be even of order 1 $\mu$Jy sr$^{-1}$, the spatially averaged density fluctuation term must be of order $\sqrt{ \braket{\delta_b^2} \braket{\delta_a^2}} \sim 10^{18}$ for a 1 eV/c$^2$ axion. This is many orders of magnitude above what current simulations and observations suggest \cite{Klypin_2018}. One can also make rough analytical arguments assuming different turbulent cascades for baryon and dark matter densities, but we find these to result in $\sqrt{ \braket{\delta_b^2} \braket{\delta_a^2}} \lesssim 10^{15}$ under the most extremely liberal of assumptions. Hence, this then gives us an estimate of the upper bound out to a line of sight of $10$ Mpc as:

\begin{align}
    L_{f, \Omega, \text{peak}} \leq 10^{-8} \text{ Jy/str} \cdot \left( \frac{m_a}{\text{eV}}\right)  \left( \frac{\mu_e}{\text{u}}\right)^{-1} 
\end{align}

\indent Recalling that the best instruments are able to observe at the $\mu$Jy level, this result gives a strong cosmological limit to ever observing the conversion of axions into photons in diffuse plasmas in the local universe, except in the most extreme of cases where our underlying assumptions have broken down. However, we expect these exceptions to be so rare, and confined to such small spaces, that we are unable to resolve them with any modern observing technologies.

\begin{figure*}
\includegraphics[width=0.9\linewidth]{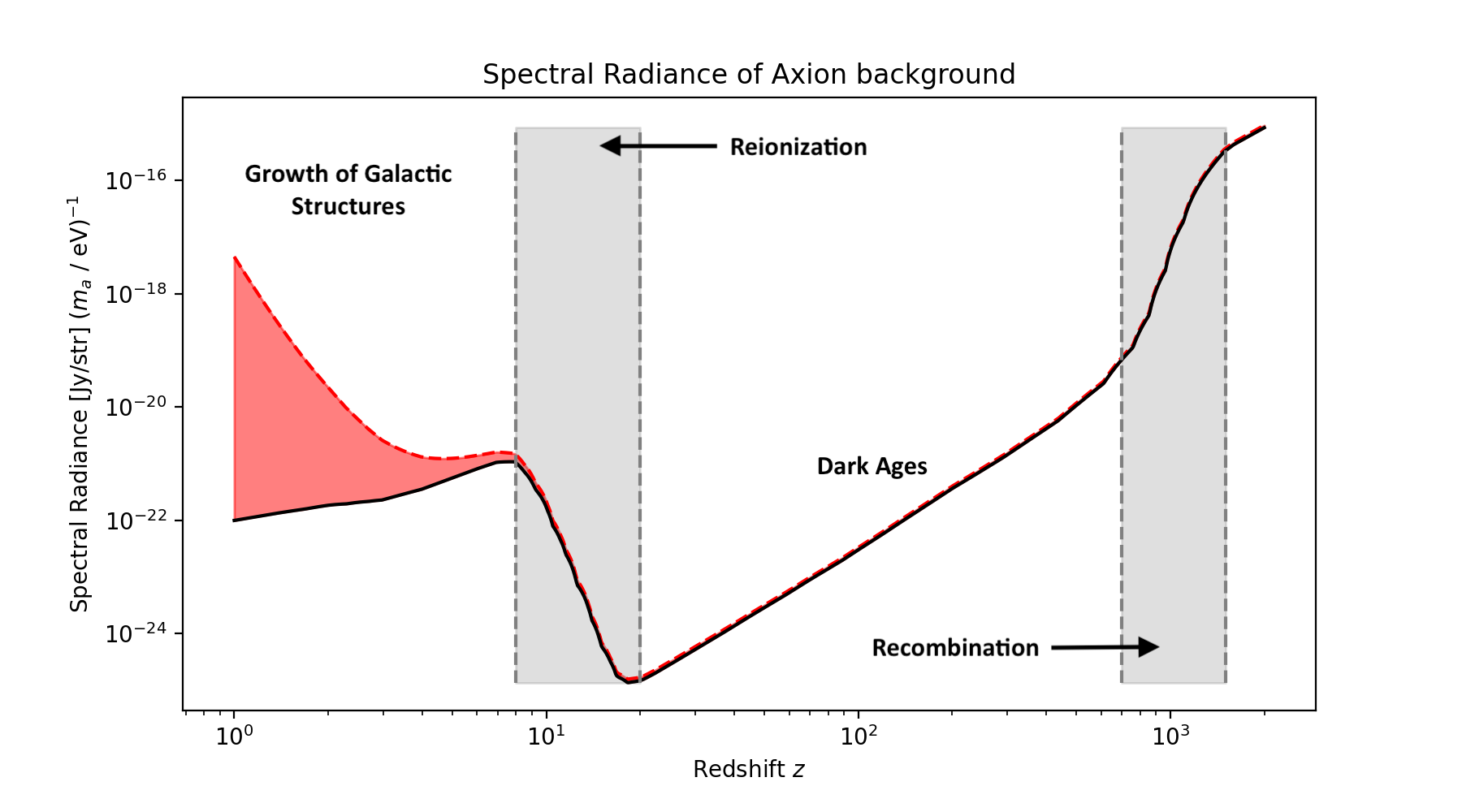}
    \caption{The computed spectral radiance of Axion-Photon conversion as a function of axion mass $m_a$ across redshift. We take $\Omega_b = 0.0224/h^2$, $\Omega_c = 0.120/h^2$, $\Omega_\Lambda = 0.685$, $\Omega_r = 2.5\times10^{-5}/h^2$, $\rho_{c,0} = 10^{-29}$ g/cm$^{-3}$, and $H_0 = 67.4$ km/s/Mpc \cite{Planck}, as well as $\mu \approx 1 $ u. The black line denotes the expected background radiance computed from matter averages, while the red region gives a rough estimate of the upper bound from increases in matter perturbations.\label{fig:axionbackground}}
\end{figure*}

\subsection{Axion-Photon Conversion from Recombination to Late Redshifts}

\indent We now consider the potential of observing a cosmological background axion signal. In this case, we can treat every volume of space as a homogeneous emitter, that shifts in frequency with the rate $\dot{f}$. To first order, one can ignore the shape of the spectral line, giving the resulting spectral energy density at a given time as $(dL/dV) /\dot{f}$. Notice, this quantity will scale as $a^{3}$. The spectral energy density at current time, ignoring any radiative transport effects, will then be:

\begin{equation}
    \begin{split}
    \frac{dE}{dVdf} = h\frac{\sigma \rho_{c,0}^2}{a^4} \frac{\Omega_{b} \Omega_{c} \mathcal{F}_e(a)}{\mu m_a }\sqrt{\frac{8 k_B T(a)}{\pi m_e}} \times \\
    \times \left[ H_0 \sqrt{\frac{\Omega_r}{ a^4}  + \frac{\Omega_b + \Omega_c}{ a^3} + \Omega_\Lambda}\right]^{-1}
    \end{split}
\end{equation}

\indent Here we use $\mathcal{F}_e(a)$ to refer to the ionization fraction at scale factor $a$, $\mu$ as the average molecular mass, and $T(a)$ as the temperature at scale factor $a$. Scale factor can also be directly related to observing frequency $f = af_a = a m_a c^2/h$. Also note that the spectral energy density can be converted to a background spectral radiance by multiplying by $c/4\pi$. For the $\Lambda$-CDM parameters, we take the 2018 Planck Collaboration values \cite{Planck}.

\indent Like in the previous section, one can also place a rough upper bound on the possible observable spectral radiance from matter contrast perturbations. One can make the ansatz of exponential growth in time for $\braket{\delta_i^2}$. Fitting for the previous result $\sqrt{\braket{\delta_b^2}\braket{\delta_a^2}} \approx 10^{15}$ at current time, and taking $\sqrt{\braket{\delta_b^2}\braket{\delta_a^2}} \approx 10^{-1}$ at recombination shows that density perturbations only help to increase the expected radiance well after reionization (Fig \ref{fig:axionbackground}). 

\indent Figure \ref{fig:axionbackground} shows the expected radiance at any redshift is well below any current telescope sensitivities. If one extends to higher redshifts, any produced photons will quickly be scattered in the ionized medium, and add to the thermal background, rendering them undetectable. One can therefore rule out the detection of axions through conversion into photons by individual charged particles in the early universe as well.

\section*{Conclusion}

\indent To summarize our results from sections \ref{sec:axionbackground} and \ref{sec:axioninism}, any detectable signal from axion-photon conversion processes will have to make use of the most extreme magnetized plasma environments in the local universe: neutron stars. Specifically, we do not expect any detection of a diffuse background signal emanating from the interstellar medium, either through direct axion / free-fermion interactions (this work) or via magnetic turbulence, in disagreement with \cite{Kelley_2017}. The former will be stronger, but still generally in the regime of $\lesssim$ 10$^{-15}$\,$m_a$\,Jy/eV for our Milky Way. Expanding to cosmological scales, we also argue that there exists a liberal upper limit of 10$^{-8}$ $m_a$ Jy/eV from this process in the local universe. In addition, we also considered a potential red-shifted  signal from recombination, yet this too falls within similar expectation of $10^{-16}$\,$m_a$\,Jy/eV. 

\indent In contrast, we have shown that axions will produce a potentially detectable background from conversion in neutron star magnetospheres. The important addendum to this result is that detection requires the use of higher order statistical methods with the background. Interestingly these methods highlight a useful relationship between population spectra and image statistics that we are expanding on in future work for the astronomical community. The underlying assumptions of our results rest on a heuristic model we have developed to model the ensemble behavior of all neutron stars in the Milky Way. There is some uncertainty in our results, as we have simplified some behavior of neutron stars in light of current uncertainties in their dynamics. Future work should update some of these simplifications, and also investigate how our results differ under contrasting dynamic models. 

\indent Our results show that a modern instrument like ALMA would be sensitive to the background in the high frequency regime (200--900\,GHz) using the high-order methods described herein. In our calculations for this, we assume a $1.5\times1.5$\,deg survey patch with 1\,min integration times, requiring a total observation time on the order of 100\,hr, pointing towards the Galactic Center, with spectral resolution on par with neutron star kick velocities ($\sim$300 km/s).

\indent At lower radio frequencies, modern instrumentation like DSA and VLA remains insensitive to this expected background from standard axion models. However, we hope that the statistical methods described will allow for new survey methodologies as described in section \ref{sec:surveymethod} that leverage background detections to find potential point-sources. On the other hand, it may be possible, with advancing technology, to detect this background at other frequencies through not-yet-existing instrumentation, specifically optimized and designed to probe the axion-pulsar background, like a simple space-based array built for a long observation campaign.

\appendix 

\section{Determining statistical moments/cumulants of confused backgrounds\label{apx:cumulants}}

\indent Consider a patch of sky within a primary antenna beam that has a well defined point-source population spectrum: $dN / dS$. The interferometric imaging of such a patch effectively sub-samples this spectrum, inducing a distribution in the background power received per `pixel' in the final image. The first and second-order moment statistics (the average power, and confusion noise level) of these pixels are oft-used in the literature, however the use of higher-order statistics is limited by the lack of a simple framework to compute such statistics. We therefore present such a framework to compute \emph{cumulant} statistics for any $dN / dS$. We choose to focus on cumulant statistics by virtue of their definition: they are additive in random variables. This eases the analysis of combined backgrounds or noise sources.

\indent We start by discretely `histograming' the point-source population in the primary beam into power bins $\Delta S_i$ with source count $dN_i = dN / dS \cdot \Delta S_i$. Each pixel will have its own histogram as well, but with point-source counts becoming random variables: $d\tilde{N}_i$. If $\Omega_p$ and $\Omega_b$ are the beam areas of the primary beam and the point-spread-function of the instrument, it is clear that we expect $\langle d\tilde{N}_i\rangle = \Omega_b/\Omega_p \cdot dN_i$. Note, when $\Omega_p$ and $\Omega_b$ are of similar order of magnitude, $d\tilde{N}_i$ will be binomially distributed, however in the high-resolution limit, when $\Omega_b << \Omega_p$, $d \tilde{N}_i$ will approach a Poisson distribution. We are also assuming that the final image is created in an area of the primary beam that is largely constant in beam response.  

\indent We will consider the high-resolution limit, where $d\tilde{N}_i \sim \text{Poiss}(\Omega_b/\Omega_p \cdot dN_i)$. Hence, the total power received in a pixel is the weighted sum of Poissonian variables, which is a non-trivial distribution. However, computing all cumulants of the distribution is made simple by the fact that the cumulants of a Poisson distribution are equal to the mean! Hence it follows, rather nicely, that the cumulants of the pixel power distribution are the scaled \emph{raw moments} of the point-source spectrum:

\begin{align}
    \kappa_n (S_\text{pix}) &= \kappa_n \left(\sum_i S_i d \tilde{N}_i \right) = \sum_i S_i^n \langle d \tilde{N}_i\rangle \\
    &= \frac{\Omega_b}{\Omega_p}\sum_i S_i^n \frac{dN}{dS} \Delta S_i \rightarrow \frac{\Omega_b}{\Omega_p} \int S^n \frac{dN}{dS}dS
\end{align}

\indent Note, the first, second, and third cumulants are respectively, the mean, variance, and central third moment. Higher order moments can be directly computed from the cumulants using Bell polynomials, hence the probability distribution of the confused background and all of its sample statistics are entirely constrained by $dN /dS$:

\begin{align*}
    \langle S_\text{pix} \rangle &= \kappa_1 & \text{Var}(\langle S_\text{pix} \rangle) &= \frac{1}{N}\kappa_2 \\
    \langle \sigma^2_\text{pix}\rangle &= \kappa_2 & \text{Var}(\langle \sigma^2_\text{pix}\rangle) &= \frac{1}{N} \left(\kappa_4 +  2 \kappa_2^2\right) \\
    \langle \mu_3 \rangle &= \kappa_3 & \text{Var}(\langle \mu_3\rangle) &= \frac{1}{N}(\mu_6 - \mu_3^2) \\
    \langle \mu_4 \rangle &= \kappa_4 + 3 k_2^2 & \text{Var}(\langle \mu_4 \rangle) &= \frac{1}{N}(\mu_8 - \mu_4^2) \\
    \vdots & & \vdots
\end{align*}
\begin{align*}
    \langle \mu_n \rangle &= \sum_{k=1}^n B_{n,k}(0,\kappa_2, ... , \kappa_{n-k+1})\\
    \text{Cov}(\langle \mu_n \rangle, \langle \mu_m \rangle) &= \frac{1}{N} (\mu_{n+m} - \mu_n \mu_m)
\end{align*}

\indent For brevity, the variance in the sample statistics at high order are given using the central moments due to the increasing complexity of Bell's polynomials. For example, for the sample variance of the third and fourth central moments, the full expression in terms of cumulants is:

\begin{align*}
    \text{Var}(\langle\mu_3\rangle) &\approx \frac{1}{N}\left( \kappa_6 + 15 \kappa_4 \kappa_2 + 9 \kappa_3^2 + 15 \kappa_2^3\right) \\
    \text{Var}(\langle\mu_4\rangle) &\approx \frac{1}{N}\big( \kappa_8 + 28 \kappa_6 \kappa_2 + 56 \kappa_5 \kappa_3 + 35 \kappa_4^2 + \\
    210 &\kappa_4 \kappa_2^2 + 280 \kappa_3^2\kappa_2 + 105 \kappa_2^4 - [\kappa_4 + 3 \kappa_2^2]^2\big)
\end{align*}

\indent The power of this framework is that we can easily combine different backgrounds and compute the significance of different statistical metrics. Take, for example, a white noise (thermal) background and some confused background characterized as a gamma distribution with parameters $(\alpha, \theta)$. The resulting cumulants are simply:

\begin{align*}
    \kappa_1 &= \alpha \theta & \kappa_2 &= \sigma_\text{th}^2 + \alpha \theta^2& \kappa_3 &= 2\alpha \theta^3 \\
    \kappa_4 &= 6\alpha\theta^4 & \kappa_5 &= 24 \alpha \theta^5 & \kappa_6 &= 120 \alpha \theta^6
\end{align*}

\indent Here the interpretation for the gamma distribution is that $\alpha$ represents the expected number of sources in a pixel, and $\theta$ represents the average power of a source – a reasonable first draft of a distribution for a confusion noise model. We can therefore see that the uncertainty in the first three sample moments are:

\begin{align*}
    \text{Var}(\langle S_\text{pix}\rangle) &= \frac{\sigma_{th}^2 + \alpha \theta^2}{N} \\
    \text{Var}(\langle\sigma^2 \rangle) &= \frac{1}{N}\left( [6\alpha + 2 \alpha^2]\theta^4 + 4\alpha  \sigma_{th}^2 \theta^2 + 2 \sigma_{th}^4\right) \\
    \text{Var}(\langle \mu_3 \rangle) &= \frac{1}{N}\big( 120 \alpha \theta^6 + 15\cdot 6 \alpha\theta^4 (\sigma_{th}^2 + \alpha \theta^2) + \\
    &9 \cdot 4\alpha^3 \theta^6 +15(\sigma_{th}^2 + \alpha \theta^2)^3\big)
\end{align*}

\indent If we want to test whether there is an extra source of noise in the system, moments can be calculated from the image, and compared with the expected moments and their variance. It follows that hypothesis testing can be easily applied on different noise models until a best-fit is found. Conversely, if we want to find a specific moment (or set of moments) that is most sensitive to some added noise, this framework allows us to determine which metrics we expect significant deviation in.

\indent For example, in the case just considered: if we have a confusion limited image ($\sigma_{th}^2 << \alpha\theta^2$), and where $\alpha$ is much bigger than one, the variance in $\mu_3$ is simply $\frac{15}{N}\alpha^3\theta^6 = \frac{15}{N}\mu_2^3$. Hence if we have a confusion noise of $0.5 \mu$Jy/beam, and 4,000,000 pixels, the standard deviation in the third moment estimate will be $1.3\times10^{-3} \mu$Jy$^3$.

\indent It also quickly becomes apparent that an algebraic implementation of this scheme is rather involved. We therefore recommend a numerical approach be used, where the relevant moments of the pixel power distribution are calculated using the cumulants calculated from $dN /dS$. Subsequent error estimates can then be quickly derived from the calculated moments. 

\indent However, we will note a useful limit. In a case where the only confounding source of noise is thermal, one can easily find the limiting thermal noise to make a significant ($q\sigma$) measurement of an expected moment $\hat{\mu}_n$ across $N$ samples:

\begin{align}
    \tilde{\sigma}_\text{th} &= \left[\frac{\hat{\mu}_n}{q} \sqrt{\frac{N}{A_n}} \right]^{1/n}  \text{where: } \\
    \nonumber \\
    A_n &= \begin{dcases*}
        \frac{(2n)!}{2^n n!} & if n is odd, \\
        \frac{1}{2^n n!} \left[(2 n)!-\frac{(n!)^3}{\left(\frac{n}{2}!\right)^2}\right] & if n is even
    \end{dcases*}
\end{align}

\indent Put more compactly, one can also show that the variance in different moment statistics in a thermally-limited image is given by:

\begin{equation}
    \text{Var}(\hat{\mu}_n) = \sigma_\text{th}^{2n} \frac{A_n}{N}
\end{equation}

\indent For more complicated noise sources, like the gamma distribution used before, one should implement a numerical scheme.

\indent As a final note, we want to introduce a generalized moment-cumulant relationship that includes real-world systematics. Generally, the observed image can be written as a linear transform $C_{ij}$ of the 'ideal' image -- e.g. as a convolution with a point-spread function. In this case the observed moments can be written in terms of the underlying cumulants, weighted by the partition moments, $M^{(n)}_{\lambda}$, of the underlying transform (Eq.~\ref{eq:contcorr}). 

\begin{equation}
    \begin{split}
    \avg{\mu_n} &= \frac{1}{N_{\text{pix}}} \sum_{\lambda \in \Lambda(n)} \frac{n!}{N(\lambda)} M^{(n)}_{\lambda} \cdot \kappa_{\lambda}(S) \\
    N(r_1 ... r_z) &= m_1! ... m_n! \cdot r_1! ... r_z! \\
    \kappa_\lambda &= \kappa_{r_1} ... \kappa_{r_z}
    \end{split} \label{eq:contcorr}
\end{equation}

\indent In the case of a convolution, these moments reduce to products of $\ell$ norms of the kernel - e.g. the PSF. Hence, higher order cumulant estimates are affected more strongly by side-lobe noise. However, if images can be composed such that the underlying PSF retains a norm of $\sim 1$ up to some $\ell-n$ norm (e.g. by increasing pixel sizes to the PSF size), the observed moment statistics will only be minimally biased. We will be releasing a future paper on this topic to discuss these nuances in more detail.

\section{Background Subtraction Error\label{app:backsub}}

\indent While there are many possible techniques and approaches to estimating a spectrally smooth foreground and background, we will consider the use of a simple convolution filter. The general approach will be to consider the radio spectrum of given at $\vec{\theta}$ and decompose it into spectrally smooth and rough components. Due to spectral lines and white noise having the same frequency (in time-delay space) structure, this is akin to creating a filter sensitive to \emph{uncorrelated} thermal noise in the radio instrument. Note, one could create a three dimensional filter across $(\vec{\theta},\nu)$, which may better remove correlated thermal noise between pixels due to the interferometric measurement process. However, for simplicity, we will stay with a filter across radio frequency.

\indent Given a search for a spectral line of width $\delta\nu$, one can find a local background strength without bias from the line through a simple moving average over a bandwidth $\Delta \nu$, with the central $\delta \nu$ section subtracted out:

\begin{align}
    S_B(\nu) &= \int \mathcal{K}(\bar{\nu})S(\nu - \bar{\nu}) d\bar{\nu}  \\
    \mathcal{K}(\bar{\nu}) &= 1/(\Delta \nu - \delta \nu) \text{ if } \delta \nu < 2|\bar{\nu}| < \Delta\nu
\end{align}

\indent The resulting transfer function of the smoothing kernel and remainder are:

\begin{align}
    H(\tau) &= \frac{1}{\sqrt{2 \pi}} \frac{\left[\Delta \nu \text{ sinc}(\Delta \nu \cdot \pi \tau) - \delta \nu \text{ sinc}(\delta \nu \cdot \pi \tau)\right]}{\Delta\nu - \delta \nu} \\
    H_r(\tau) & = \frac{1}{\sqrt{2\pi}} - H(\tau)
\end{align}

\indent One can show that the transmission loss of a spectral line signal of width $\delta \nu$ through this filter will be on the order of $<10^{-2} \text{ dB}$ for any naive set up of this filter with $\Delta \nu > 10 \delta \nu$. This leaves the intended signal of interest still intact, but what remains of the smooth background in the remainder? This will be a function of the expected power spectrum of the background. However, as we expect it to be smooth, we expect the power spectrum to be steep in $\tau$. In principle, $H_r(\tau)$ should be able to remove such red-noise, but it becomes a strongly context-dependent problem to solve for exactly. 

\indent However, if one considers a simple situation where we have 1000 samples sampled linearly in frequency, and apply a filter (in bin sizes) with $\delta \nu = 1$, $\Delta \nu = 10$, any red noise process with a spectral index $\gamma>0.3$ will be attenuated by at least 10 dB. In fact, for the spectral indices, $\{1,2,3,4\}$, we have the following attenuation: $(-20, -55, -70, -75)$ dB. Since most models of background confusion sources show an index around $\gamma \sim 1$, we should expect noise from the smooth background to be reduced by at least two orders of magnitude in the remainder image.

\indent In practice, a matched filter should be designed using prior knowledge of the smooth background statistics to maximize the SNR of the target spectral line signal. This will further reduce the subtraction noise in the remainder images used to search for spectral lines. Further, since images will already be thermally limited, this leaves thermal noise as the primary source of uncertainty in moment estimates of the remaining background. In the case of extremely strong point-sources in the foreground, care should be taken to identify and clean them from the images. 

\section{Telescope Survey Parameters \label{apx:instinfo}}

\begin{table*}[t]
    \centering
    \begin{tabular}{|c|c|c|c|c|c|c|}
        \hline Instrument & Frequences [GHz] & $A_e$ [m$^2$] & $T_\text{sys}$ [K] & $N_\text{ant}$ & $\sqrt{\Omega_b}(\nu_\text{ref})$ [arcsec] & $\nu_\text{ref}$ [GHz] \\ \hline
        DSA & $[0.8 -2.1]$ & 28.3 & 17 & 1650 & 3.3 & 1.35  \\
        VLA & $[0.1-50]$ & 490 & 40 & 28 & 23 & 3 \\
        HIFI & $[450-2000]$ & 9.6 & 3 & 1 & 14 & 2000\\
        ALMA & $[50-950]$ & 113 & 100 & 66 & 0.5 & 950 \\ \hline
    \end{tabular}
    \caption{The relevant instrument parameters used in calculating axion-background sensitivities for figures \ref{fig:AxionSens} and \ref{fig:AxionSens_Realistic}.}
    \label{tab:InstrumentInfo}
\end{table*}

\indent Table \ref{tab:InstrumentInfo} lists the relevant instrument information used for the axion sensitivity figures \ref{fig:AxionSens} and \ref{fig:AxionSens_Realistic}. Some of these values, like $T_\text{sys}$, are approximate and can vary across radio frequency for a given instrument. $\Omega_b$, varies dramatically with radio frequency, so we use equation \ref{eq:beamscaling} to adjust the synthesized beam area for different radio frequencies.

\begin{equation}
    \Omega_b(\nu) = \Omega_b(\nu_\text{ref}) \cdot \left[\frac{\nu}{\nu_\text{ref}}\right]^{-2} \label{eq:beamscaling}
\end{equation} \\

\section{Synthetic Pulsar Detections with Existing Surveys} \label{app:APPLPy}

\indent To validate our neutron star model, we put together a new package, \texttt{APPLPy} (A Pulsar Population Library in Python), that contains a record of all modern pulsar surveys in the ATNF pulsar catalog and can compute if a given pulsar could have been detected by any such surveys. We used a simple model that corrects for both dispersion-measure (DM) smearing and long period red noise in pulsar SNR calculations. For simplicity we assign pulsars a DM according to $r \cdot 3 \times 10^{-2}$ cm$^{-3}$, give each pulsar an intrinsic duty cycle of 0.2, and give all pulsars a beaming fraction of 15$\%$ While these are simplifications, we believe they are accurate enough to give reasonable results.

\indent More information and exact details on the surveys and SNR models used can be found in the documentation for \texttt{APPLPy}, which can be found here: \url{https://github.com/FelixWeber02/APPLPy}.


\bibliography{refs2}

\end{document}